\documentclass[journal,10pt]{IEEEtran}

\usepackage{amsmath}

\usepackage[cmintegrals]{newtxmath}

\usepackage{amsmath}
\usepackage{graphicx}
\usepackage{epstopdf}
\usepackage{wrapfig}
\usepackage{float}
\usepackage{cite}
\usepackage{stfloats}
\usepackage{multirow}
\usepackage{array}
\usepackage{booktabs}
\usepackage{longtable}

\begin{document}

\title{Performance Analyses of TAS/Alamouti-MRC NOMA in Dual-Hop Full-Duplex AF Relaying Networks}

\author{Mesut Toka, Eray G\"{u}ven, Mehmet Akif Durmaz, G\"{u}ne\c{s} Karabulut Kurt, O\u{g}uz Kucur

\thanks{This work was supported by the Scientific and Technological Research Council of Turkey (T\"{U}B\.{I}TAK) under Grant EEEAG 118E274.}
\thanks{Mesut Toka and O\u{g}uz Kucur are with the Department of Electronics Engineering, Gebze Technical University, Gebze/Kocaeli, 41400, Turkey (e-mail: tokamesut@gmail.com, okucur@gtu.edu.tr). Mesut Toka is also with the Department of Electrical and Electronics Engineering, Ni\u{g}de \"{O}mer Halisdemir University, Ni\u{g}de, 51240, Turkey.}
\thanks{Eray G\"{u}ven, Mehmet Akif Durmaz and G\"{u}ne\c{s} Karabulut Kurt are with the Department of Electronics and Communications Engineering, \.{I}stanbul Technical University, \.{I}stanbul, 34469, Turkey (e-mail: \{guvenera, durmazm, gkurt\}@itu.edu.tr).}}

\markboth{IEEE Transactions on XXX,~Vol.~XX, No.~XX, XXX~2020}
{}

\maketitle

\begin{abstract}
In this paper, the performance of a power domain downlink multiple-input multiple-output non-orthogonal multiple access system in dual-hop full-duplex (FD) relaying networks is investigated over Nakagami-$m$ fading channels by considering the channel estimation error and feedback delay. Particularly, in the investigated system, the base station equipped with multiple antennas transmits information to all mobile users by applying conventional transmit antenna selection/Alamouti-space-time block coding scheme with the help of a dedicated FD amplify-and-forward relay. The received signals at mobile users are combined according to maximal-ratio combining technique to exploit benefits of receive diversity. In order to demonstrate the superiority of the proposed system, outage probability (OP) is investigated and tight lower bound expressions are derived for the obtained OP. Moreover, asymptotic analyses are also conducted for ideal and practical conditions to provide further insights about the outage behavior in the high signal-to-noise ratio region. Finally, theoretical analyses are validated via Monte Carlo simulations and software defined radio based test-bed implementation. 	        
\end{abstract}


\IEEEpeerreviewmaketitle

\section{Introduction}  
\IEEEPARstart{R}{ecently}, non-orthogonal multiple access (NOMA) has been regarded as an efficient technique due to the weakness of currently utilized orthogonal multiple access (OMA) techniques to fulfill the requirements of high quality services demanded by the unprecedented forthcoming communication applications \cite{LiuY,Aldababsa}. Power-domain NOMA, which is based on allowing multiple users with different power levels determined according to channel statistics to utilize the same time/frequency resource simultaneously, is the most preferred one among all NOMA techniques due to its simple applicability to existing communication networks. At the receiver side, corresponding users apply successive interference cancellation (SIC) techniques to decode their own information \cite{Saito1,Islam}. Hence, NOMA has been extensively investigated in the literature due to its inherent ability of increasing spectral efficiency and ensuring massive connectivity together with fairness among users \cite{DingZ,ZhengY}. In \cite{DingZ}, a downlink NOMA system based on fixed power allocation assumption, in which users' distances from base station (BS) are distributed uniformly in a cell-cluster, is investigated through ergodic rate and outage probability (OP). Also, the same NOMA network in \cite{DingZ} is investigated by taking into account channel estimation errors (CEEs) in \cite{ZhengY} and simple closed-form approximations for OP and average sum rate are derived. On the other hand, it is obvious from OMA literature that multi-antenna and cooperative techniques provide considerable enhancement on the performance and coverage area of wireless networks \cite{Foschini,Laneman2}. Therefore, there are many works in the literature that have attempted to combine multiple-input multiple-output (MIMO) and cooperative techniques with NOMA to exploit benefits of spatial diversity and improve transmission reliability, respectively. In this context, our attention is focused on investigating MIMO-NOMA structure in cooperative networks.   

\subsection{Literature Review}
In \cite{DingZ1}, the authors have focused on design of precoding/detection structure for MIMO-NOMA system and investigated the impact of user pairing scheme to enlarge the performance gap as compared to the counterpart OMA. However, exploiting precoding based techniques has caused ideal channel state information (CSI) burden in wireless communications systems. On the contrary, space-time block coding (STBC) \cite{Alamouti} does not require CSI, so it is more robust to CEEs when compared to the other MIMO techniques. Therefore, in \cite{ChoiJ}, a downlink NOMA system, in which two coordinated BSs with single antenna perform distributed Alamouti-STBC to increase the reliability of cell-edge user, has been proposed and investigated through transmission rates. In \cite{Mtoka}, conventional Alamouti-STBC scheme has been applied to downlink multi-user NOMA system, where users are equipped with single antenna, and exact OP expression is obtained for any user over Nakagami-$m$ fading channels. In \cite{MtokaTVT}, the same system investigated in \cite{Mtoka} has been generalized to cover all STBC schemes \cite{Tarokh} by taking into account practical impairments such as CEE, feedback delay (FBD) and imperfect SIC. However, using multiple antennas at the transmitter and receiver ends increases hardware complexity and power consumption due to multiple radio-frequency (RF) chains. In order to overcome these drawbacks, antenna selection (AS) approach has been considered since it reduces the number of used RF chains as well as retaining many advantages of spatial diversity. Also, only partial CSI is required within the scope of AS. Thereby, several optimal and sub-optimal AS techniques have been proposed and investigated for OMA systems so far. In addition, some sub-optimal AS techniques such as max-max-max, max-min-max and maximum-channel-gain-based one have been addressed for MIMO-NOMA systems consisting of one BS and two users in \cite{YYu1}. In \cite{QQLi}, by considering the similar system in \cite{YYu1}, the authors have proposed a joint AS scheme, which ensures minimizing the OP by selecting the best antenna at the BS and optimal antennas at the two users, and obtained closed-form OP for Nakagami-$m$ fading channels. However, previously proposed AS schemes consider selection processes based on individual users. Therefore, the authors of \cite{AldababsaMajTAS} have proposed a novel AS scheme, which selects the best antenna of BS according to the majority decision of users, for single hop multi-user NOMA and investigated the OP performance over Nakagami-$m$ channels in the presence of CEE and FBD effects. In \cite{MTokaTETT}, by using the AS technique proposed in \cite{AldababsaMajTAS}, transmit AS (TAS)/Alamouti-STBC scheme is investigated for single-hop MIMO-NOMA with practical impairments.

Besides multi-antenna schemes, there are many studies considering relaying techniques in NOMA network with/without MIMO in the literature due to the demands on expanding the coverage area. Accordingly, in \cite{JBKim}, a downlink NOMA network based on coordinated direct and relay transmission, where the BS directly communicates with one user while a dedicated relay assists the other user, has been investigated. In \cite{JJmen2}, performance of a downlink cooperative NOMA network in which the BS serves multiple users through a dedicated amplify-and-forward (AF) relay has been analyzed in terms of OP over Nakagami-$m$ fading channels by also considering CEE effects. In \cite{Kader}, Alamouti-STBC scheme has been investigated in a cooperative decode-and-forward (DF) relaying network with direct link, where the BS communicates with only one user by adopting NOMA principle. On the other hand, AS techniques have also been investigated in cooperative NOMA networks. In \cite{ZhangY}, OP performance of a downlink cooperative NOMA network with AF relay, where the BS and multiple users apply TAS and maximal-ratio combining (MRC) techniques, respectively, has been analyzed over Nakagami-$m$ fading channels under CEE effects. In \cite{AldababsaTASMRC}, the same system in \cite{ZhangY} has been addressed in case of not only CEE but also imperfect SIC effects when the relay operates in variable and fixed gain modes. In \cite{MtokaSIU}, outage performance of TAS/Alamouti scheme is studied in dual-hop NOMA network for HD AF relay and two users by only simulations over Rayleigh fading channels.  

Aforementioned studies within the scope of cooperative NOMA networks are based on half-duplex (HD) relaying technique. Although HD relaying increases reliability of the system, it reduces the spectral efficiency due to its principle of allocation of two orthogonal channels. Fortunately, to overcome this problem, full-duplex (FD) relaying has been proposed and regarded as a promising technique due to its ability of improving the spectral efficiency \cite{Duarte,Rodriguez,MtokaIET}. Since reliability of the system becomes crucial when the direct link between the BS and users does not exist due to huge obstacles or heavy shadowing, some studies have focused on exploiting benefits of the dedicated FD relays within the scope of cooperative NOMA \cite{XYue,ZhongC,MFKader,TMCC,MMohammadi,Tregancini}. In \cite{XYue}, the authors propose and investigate a cooperative two-user NOMA network with/without direct link, where one user close to the BS acts as a DF relay switching between FD and HD mode to assist the other user far to the BS. To measure the level of performance, OP, ergodic rate and energy efficiency have been addressed in the case of Rayleigh fading channels. In \cite{ZhongC}, a cooperative two-user NOMA network with DF FD relay has been proposed and OP performance together with ergodic sum capacity have been investigated over Rayleigh fading channels. Particularly, in the investigated system, the BS transmits information directly to the relay and near user while the relay forwards decoded information to far and near users at the same time. In \cite{MFKader}, a cooperative three-user NOMA network, where the BS communicates with two far users through a DF FD relay, has been considered and analyzed in terms of OP and ergodic sum capacity. In \cite{TMCC}, exactly the same system investigated in \cite{ZhongC} has been considered and analyzed over independent and identically distributed (i.i.d.) Nakagami-$m$ fading channels in terms of OP together with the ergodic rate, however the residual self-interference (SI) at the FD relay is assumed as Gaussian distribution. In \cite{MMohammadi}, the same system in \cite{ZhongC} has been addressed in point of AS problem by assuming that the BS and relay are equipped with multiple antennas. For performance criterion, closed-form OP and ergodic sum rate expressions are obtained. In addition, relay selection techniques have been investigated in two-user cooperative NOMA with multiple AF FD relays over Rayleigh fading channels in \cite{Tregancini}. However, in \cite{Tregancini}, exact closed-form OP expressions could not be provided, and the analyses have been supported through lower bound and asymptotic approximations. 

\subsection{Motivation and Contributions}
The studies on FD-NOMA network mentioned above in the literature review have demonstrated that the spectral efficiency of a cooperative NOMA system can be increased via FD relaying and also FD-NOMA outperforms the HD-NOMA counterpart under some certain conditions. In addition, using multiple antennas in NOMA has been shown to be quite efficient in improving the system performance. Moreover, given the significant advantages of TAS and Alamouti-STBC multiple antenna techniques mentioned in the literature review such as a reduced CSI burden together with benefits of spatial diversity and reduced RF complexity and power consumption, we believe that it is reasonably exciting to exploit the combination of them jointly. With the motivation of promising results gained from MIMO-NOMA and cooperative NOMA networks examined above, in this paper, our main focus has become investigating conventional TAS/Alamouti-STBC scheme in dual-hop AF FD multi-user power-domain NOMA network over i.i.d. Nakagami-$m$ fading channels in the presence of CEE and FBD effects. Particularly, in the considered system, the best two antennas at the BS are selected according to conventional TAS technique to apply Alamouti-STBC scheme while receivers of mobile users are assumed to combine received information by using MRC technique. The major contributions of this paper are highlighted as follows:
\begin{itemize}
	\item Unlike most of the existing studies carried out within cooperative FD-NOMA, we consider that the FD relay operates in AF mode and also residual SI link between antennas of the relay is exposed to fading effects. In addition, this paper is the first attempt that analyzes TAS/Alamouti scheme in cooperative NOMA literature and even assumes all channels undergo Nakagami-$m$ fading. It is worth noting that benefits of both the reduced RF chains and STBC have been brought to cooperative NOMA systems with the combination of TAS and Alamouti schemes.   
\end{itemize} \vspace{-1 mm}
\begin{itemize}
	\item In order to demonstrate the level of system performance, exact OP expression for any user is derived in single-fold integral through moment generating function (MGF) approach and simple lower bounds together with asymptotic expressions are also obtained to provide more meaningful insights into the OP. Thus, with the asymptotic analyses, the impacts of CEE, FBD and residual SI on the performance in asymptotic regime are provided more clearly.    
\end{itemize} \vspace{-1 mm}   
\begin{itemize}
	\item Furthermore, test-bed implementation of the investigated system, which is the most sophisticated NOMA implementation in the literature, is conducted through USRP software defined radios (SDRs) to demonstrate its feasibility in a practical manner.  
\end{itemize} \vspace{-1 mm}   
\begin{itemize}
	\item Numerical results of the investigated TAS/Alamouti-STBC FD-NOMA system are verified by Monte Carlo simulations and SDR-based real-time tests and compared to both HD-NOMA and FD-OMA counterparts. Results demonstrate that the quality of SI cancellation in the FD relay is mostly effective in OP performances of the users with lower power levels whose performances are also highly effected by the number of antennas at the BS.     
\end{itemize} \vspace{-1 mm}          

\subsection{Notations and Organization}
Bold lowercase letter and $\|\cdot\|_{F}$ denote vectors and the Frobenius norm of a vector, respectively. $\mathbb{CN}(0,\sigma^2)$ indicates the complex Gaussian distribution with zero mean and variance of $\sigma^2$. While $E\left[ \cdot\right] $ represents the expectation operator, $Pr(\cdot)$ denotes the probability of any event.

The reminder of the paper is given as follows. The investigated system and channel statistics are introduced in Section II. The exact OP together with lower bound and asymptotic analyses are performed in Section III. Test-bed implementation of the investigated system is presented in Section IV. Finally, Sections V and VI provide numerical results and conclusions, respectively.

\begin{figure*}[!t]
	\centering
	\includegraphics[scale=0.80]{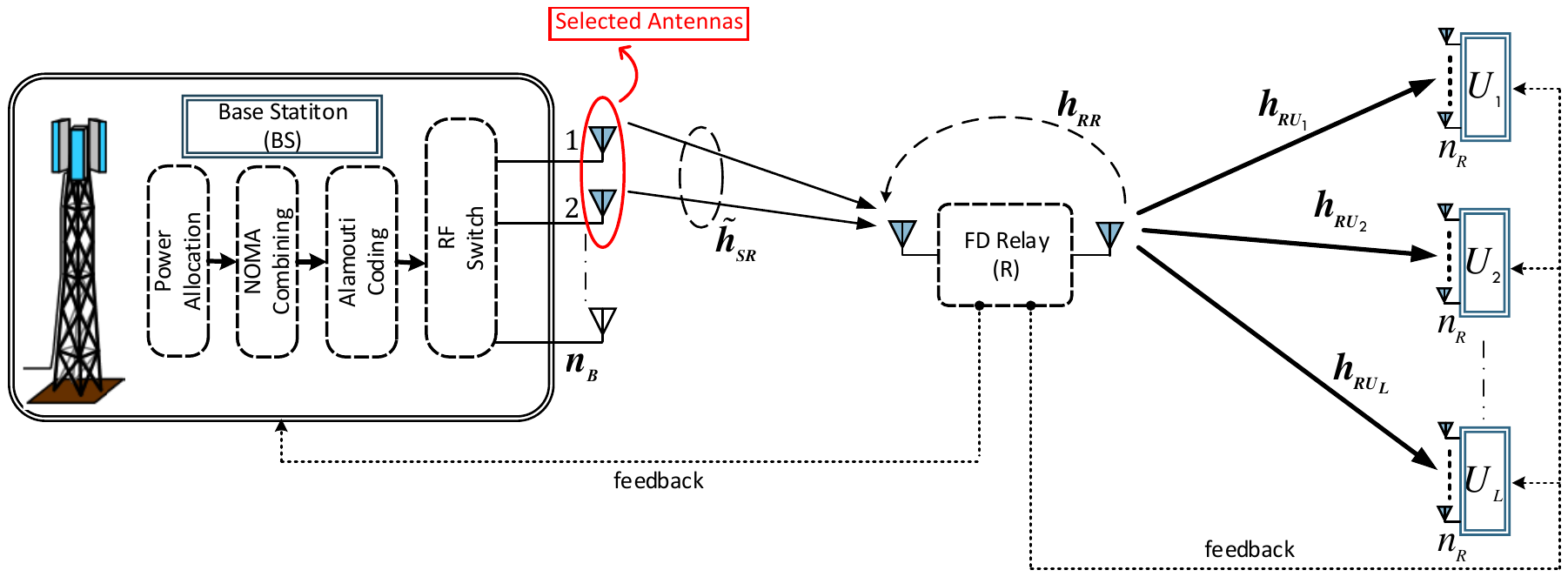}
	\setlength{\belowcaptionskip}{-2pt}
	\caption{System model}
	\label{fig:1}       
\end{figure*}

\section{System Model}
Consider a power-domain multi-user downlink MIMO-NOMA cooperative system subjected to clustered network topology, where the BS and relay are in the cluster while mobile users are out of the cell coverage as given in Fig. 1. While the BS ($S$) and mobile users ($U_l$, $l=1,2,\cdots,L$) are equipped with $n_B$ and $n_R$ antennas, respectively, the FD relay ($R$) adopting AF protocol has two simultaneously operating antennas, one for receiving and the other for transmitting. In the first hop, the BS applies conventional (not distributed) Alamouti-STBC scheme for broadcasting. Particularly, two transmit antennas of the BS providing the best signal-to-noise ratios (SNRs) belonging to $S-R$ links are selected by the relay and their indices are sent to the BS through a feedback channel. On the other hand, in the second hop, the relay forwards received signals from the BS to all mobile users whose receivers apply MRC technique to utilize the benefits of receive diversity. Within the time-slot $n$, the BS generates code matrix $\mathbf{G}_2[n]=\left\lbrace x_{tk} \right\rbrace_{2\times 2}$ ($t\in\left\lbrace 1,2 \right\rbrace $; time interval of the Alamouti-STBC code, $k\in\left\lbrace 1,2 \right\rbrace $; the index of selected antennas at the BS for Alamouti-STBC code) whose entries consist of superposed NOMA signals and their complex conjugates \cite{Alamouti,Mtoka}.

\subsection{Channel Statistics}
In the network, channel coefficient vectors related to $S-R$, $R-U_l$ and $R-R$ links are denoted by $\mathbf{h}_{SR}=\left\lbrace h_{SR}^i \right\rbrace_{n_B\times1} $, $\mathbf{h}_{RU_l}=\left\lbrace h_{RU_l}^j \right\rbrace_{1\times n_R} $ and $\mathbf{h}_{RR}=\left\lbrace h_{RR}^t \right\rbrace_{2\times1} $ ($1\leq i\leq n_B$ and $1\leq j\leq n_R$) whose elements are modeled as i.i.d. Nakagami-$m$ distribution with powers $\Omega_{SR}=E\left[ |h^i_{SR}|^2\right] =d_{SR}^{-\eta}$, $\Omega_{RU_l}=E\left[ |h^j_{RU_l}|^2\right] =d_{RU_l}^{-\eta}$ and $\Omega_{RR}=E\left[ |h^t_{RR}|^2\right] =\alpha P_R^{\mu-1}$, respectively. Here, $d_{SR}$, $d_{RU_l}$ and $\eta$ represent normalized distances of $S-R$ and $R-U_l$ links and path loss exponent, respectively. Also, we assume that the relay suffers from residual SI effect resulting from active and/or passive SI cancellation techniques which are modeled as in \cite{Duarte} and \cite{Rodriguez}. Therefore, constants $\alpha$ and $\mu$ denote the quality of SI cancellation process ($0\leq\mu\leq1$, $\alpha>0$).

In the training period before the actual transmission, the relay is assumed to estimate channel gains of both $S-R$ and $R-U_l$ links for the determination of indices belonging to be selected antennas of the BS and power coefficients to be allocated to users, respectively. In this context, we consider that CEE and FBD exist in both hops since they are inevitable in case of practical implementations. Note that there is no CEE and/or FBD effects on SI link since channel estimation process is not carried out and only SI cancellation techniques are used. In the first hop, since two antennas of the BS are selected according to optimum TAS technique at the relay, squares of the estimated channel gains are sorted in descending order as $|\hat{h}^1_{SR}|^2>|\hat{h}^2_{SR}|^2>\cdots>|\hat{h}^{n_B}_{SR}|^2$. Then, two antennas provide the highest channel gains are selected to maintain the actual transmission. $S-R$ channel coefficient vector resulting from antenna selection process is denoted by $\hat{\tilde{\mathbf{h}}}_{SR}=\begin{bmatrix}\hat{h}^1_{SR} \\ \hat{h}^2_{SR} \end{bmatrix}$. By taking into account the CEE and FBD effects, the channel coefficient vector corresponding to the first hop is re-expressed as $\mathbf{\tilde{h}}_{SR}=\rho_{SR}\hat{\tilde{\textbf{h}}}^{\tau}_{SR}+\boldsymbol{\varepsilon}_{fbd,SR}+\boldsymbol{\varepsilon}_{est,SR}$, where $\hat{\mathbf{h}}$ and $\hat{\mathbf{h}}^{\tau}$ denote the erroneously estimated channel vector and its delayed version. $\boldsymbol{\varepsilon}_{est,SR}\sim\mathbb{CN}(0,\sigma_{est,SR}^2)$ and $\boldsymbol{\varepsilon}_{fbd,SR}\sim\mathbb{CN}(0,\sigma_{fbd,SR}^2)$ denote CEE and FBD vectors whose entries' variances are found by $\sigma_{est,SR}^2=\Omega_{SR}-\hat{\Omega}_{SR}$ and $\sigma_{fbd,SR}^2=\left(1-\rho_{SR}^2\right)\hat{\Omega}_{SR}$, respectively. $\rho_X=J_0\left(2\pi f_D\tau \right) \left(0<\rho<1 \right)$ is the correlation coefficient corresponding to link $X$ where $J_0(\cdot)$, $f_D$ and $\tau$ represent zero-order Bessel function of the first kind, maximum Doppler frequency and time delay, respectively \cite{NSKim,ENO}. Then, CEE and FBD vectors can be represented by a new random variable as $\boldsymbol{\varepsilon}_{SR}=\boldsymbol{\varepsilon}_{fbd,SR}+\boldsymbol{\varepsilon}_{est,SR}$ with a variance of $\sigma_{SR}^2=\sigma_{fbd,SR}^2+\sigma_{est,SR}^2$ \cite{ZhouS,Coskun}. On the other hand, in the second hop, the channel coefficient vector of $R-U_l$ link is expressed as $\mathbf{h}_{RU_l}=\rho_{RU_l}\hat{\mathbf{h}}^{\tau}_{RU_l}+\boldsymbol{\varepsilon}_{fbd,l}+\boldsymbol{\varepsilon}_{est,l}$. Similar to the first hop, $\boldsymbol{\varepsilon}_{est,l}\sim\mathbb{CN}(0,\sigma_{est,l}^2)$ and $\boldsymbol{\varepsilon}_{fbd,l}\sim\mathbb{CN}(0,\sigma_{fbd,l}^2)$ denote CEE and FBD vectors with variances of $\sigma_{est,l}^2=\Omega_{RU_l}-\hat{\Omega}_{RU_l}$ and $\sigma_{fbd,l}^2=\left(1-\rho_{RU_l}^2\right)\hat{\Omega}_{RU_l}$. Afterwards, CEE and FBD vectors can be represented by $\boldsymbol{\varepsilon}_{RU_l}=\boldsymbol{\varepsilon}_{fbd,l}+\boldsymbol{\varepsilon}_{est,l}$ with a variance of $\sigma_{RU_l}^2=\sigma_{fbd,l}^2+\sigma_{est,l}^2$.

\section{Performance Analysis}
In this section, fundamental mathematical equations related to transmission process at the relay and any mobile user are defined and then end-to-end ($e2e$) signal-to-interference plus noise ratio (SINR) expression is derived. Then, the exact OP for any user is obtained through MGF approach. Finally, simple lower bounds and asymptotic expressions are derived to provide more meaningful insights into the system performance.

\subsection{Derivation of $e2e$ SINR}
In the first hop, the signal received by the relay in the $n$th time interval is expressed as
\begin{equation} \label{eq:1}
\mathbf{y}_R\left[ n\right]=\mathbf{G}_2\left[n \right]\mathbf{\tilde{h}}_{SR}+\mathbf{h}_{RR}\bullet \mathbf{s}_R[n]+\mathbf{w}_R[n]
\end{equation}
where $\mathbf{w}_R[n]=\left\lbrace w_R^t \right\rbrace_{2\times 1} $ denotes noise vector at the relay whose entries are distributed as $\mathbb{CN}(0,\sigma_R^2)$. While $\mathbf{s}_R[n]=\mathcal{G} \mathbf{y}_R[n-D]=\left\lbrace \mathcal{G} y_{R}^t \right\rbrace_{2\times 1} $ represents the signal vector transmitted from $R$ to $U_l$, $D$ and $\mathcal{G}$ denote processing delay and amplifying factor of the FD relay. Also, $(\bullet)$ is the component-wise product of two vectors with the same dimension. Then, the received signal vector $\mathbf{y}_{SRU_l}\left[ n\right]=\left\lbrace y_{SRU_l}^{t,j} \right\rbrace_{2\times n_R} $ by the $l$th user can be written as
\begin{equation} \label{eq:2}
\begin{split}
\mathbf{y}_{SRU_l}\left[ n\right]&=\mathbf{s}_R[n]\mathbf{h}_{RU_l}+\mathbf{w}_l[n] \\
&=\mathcal{G}\left(\mathbf{G}_2[n-D]\mathbf{\tilde{h}}_{SR}+\mathbf{h}_{RR}\bullet \mathbf{s}_R[n-D]\right.   \\
&\left.  +\mathbf{w}_R[n-D]\right)\mathbf{h}_{RU_l}+\mathbf{w}_l[n] 
\end{split}
\end{equation}
where $\mathbf{w}_l[n]=\left\lbrace w_l^{t,j} \right\rbrace_{2\times n_R} $ denotes noise vector at the $l$th user whose entries are distributed as $\mathbb{CN}(0,\sigma_l^2)$. Without loss of generality, $\sigma_R^2=\sigma_l^2=\sigma^2$ is assumed to be mathematically tractable. By substituting $\mathbf{s}_R[n]$ into (\ref{eq:2}), the received signal vector at the $l$th user can be rewritten as
\begin{equation} \label{eq:3}
\begin{split}
&\mathbf{y}_{SRU_l}[n]=\mathcal{G}\mathbf{G}_2[n-D]\rho_{SR}\rho_{RU_l}\hat{\tilde{\mathbf{h}}}^{\tau}_{SR}\hat{\mathbf{h}}^{\tau}_{RU_l} \\
&+\mathcal{G}\mathbf{G}_2[n-D]\rho_{SR}\hat{\tilde{\mathbf{h}}}^{\tau}_{SR}\boldsymbol{\varepsilon}_{RU_l}+\mathcal{G}\mathbf{G}_2[n-D]\rho_{RU_l}\hat{\mathbf{h}}^{\tau}_{RU_l}\boldsymbol{\varepsilon}_{SR} \\
&+\mathcal{G}\mathbf{G}_2[n-D]\boldsymbol{\varepsilon}_{SR}\boldsymbol{\varepsilon}_{RU_l}+\mathcal{G}\mathbf{s}_R[n-D]\rho_{RU_l}\mathbf{h}_{RR}\hat{\mathbf{h}}^{\tau}_{RU_l} \\
&+\mathcal{G}\mathbf{s}_R[n-D]\mathbf{h}_{RR}\boldsymbol{\varepsilon}_{RU_l}+\mathcal{G}\mathbf{w}_R[n-D]\rho_{RU_l}\hat{\mathbf{h}}^{\tau}_{RU_l} \\
&+\mathcal{G}\mathbf{w}_R[n-D]\boldsymbol{\varepsilon}_{RU_l}+\mathbf{w}_l[n] 
\end{split}
\end{equation}    
In (\ref{eq:3}), the first line comprises both the desired signal related to $U_l$ and interference signals coming from other users. Then, received signals by mobile users are combined according to MRC technique. With the help of (\ref{eq:1}), amplifying factor of the FD relay can be determined as 
\begin{equation} \label{eq:4}
\mathcal{G}=\sqrt{\dfrac{P_R}{P_S\rho_{SR}^2\|\hat{\tilde{\mathbf{h}}}^{\tau}_{SR}\|_F^2+P_R|h_{RR}|^2+P_S\sigma_{SR}^2+\sigma^2}}
\end{equation}
where $P_S$ is the transmit power of the BS. However, for mathematical simplicity, $P_S=P_R=P$ is assumed in the continuation of the analysis. 

Without loss of generality, we assume that power allocation coefficients ($a_l$) of mobile users are determined according to statistics of $R-U_l$ links since direct link between the BS and mobile users can not be established due to the poor channel conditions. Therefore, in order for the BS to determine the power allocation coefficients of users, the relay estimates the effective channel gains of $R-U_l$ link and sort as $\|\hat{\mathbf{h}}_{RU_1}\|_F^2\leq\|\hat{\mathbf{h}}_{RU_2}\|_F^2$ $\cdots\leq\|\hat{\mathbf{h}}_{RU_L}\|_F^2$. Then, the relay transmits the ordering to both BS and users through feedback channel. The BS determines power levels as $a_1>a_2>\cdots>a_L$ (subjected to $\sum_{l=1}^{L}a_l=1$) with the help of the ordering. By using the decoding method of STBC as in \cite{ChenS} and considering SIC technique is perfectly carried out at mobile users, instantaneous SINR that $U_l$ can decode the signal of $U_k$ ($k\leq l$) is expressed as
\begin{equation}\label{eq:5}
\begin{split}
&\gamma_{SRU_{k\rightarrow l}} \\
&=\dfrac{\frac{\bar{\gamma}^2}{2}ABa_k}{\frac{\bar{\gamma}^2}{2}AB\overset{L}{\underset{t=k+1}{\sum}}a_t+\vartheta_1\bar{\gamma}A+\vartheta_2\bar{\gamma}B+\vartheta_3\bar{\gamma}C+\vartheta_4\bar{\gamma}^2BC+\vartheta_5}
\end{split}
\end{equation} 
where $\bar{\gamma}=P/\sigma^2$ is the average SNR, and $A\stackrel{\triangle}{=}\|\hat{\tilde{\mathbf{h}}}^{\tau}_{SR}\|_F^2$, $B\stackrel{\triangle}{=}\|\hat{\mathbf{h}}^{\tau}_{RU_l}\|_F^2$ and $C\stackrel{\triangle}{=}|h_{RR}|^2$ definitions are used for simplicity. In addition, constant variables represented by notation $\vartheta$ are given below:
\begin{equation}\label{eq:6}
\begin{split}
\vartheta_1&\stackrel{\triangle}{=}\dfrac{\bar{\gamma}}{2}\dfrac{\sigma_{RU_l}^2}{\rho_{RU_l}^2}+\dfrac{1}{\rho_{RU_l}^2}~~~~~~~~~~,~~\vartheta_2\stackrel{\triangle}{=}\dfrac{\bar{\gamma}}{2}\dfrac{\sigma_{SR}^2}{\rho_{SR}^2}+\dfrac{1}{\rho_{SR}^2} \\
\vartheta_3&\stackrel{\triangle}{=}\bar{\gamma}\dfrac{\sigma_{RU_l}^2}{\rho_{SR}^2\rho_{RU_l}^2}+\dfrac{1}{\rho_{SR}^2\rho_{RU_l}^2}~~,~~\vartheta_4\stackrel{\triangle}{=}\dfrac{1}{\rho_{SR}^2}  \\
\vartheta_5&\stackrel{\triangle}{=}\dfrac{1}{\rho_{SR}^2\rho_{RU_l}^2}\left(\dfrac{\bar{\gamma}^2}{2}\sigma_{SR}^2\sigma_{RU_l}^2+\bar{\gamma}\sigma_{RU_l}^2+\bar{\gamma}\sigma_{SR}^2+1 \right) 
\end{split}
\end{equation}

\subsection{Outage Probability Analysis}
By assuming $\left\lbrace \gamma_{SRU_{k\rightarrow l}}>\gamma_{th,k} \right\rbrace $ as the event that $U_l$ can detect the message corresponding to itself or $U_k$ ($1\leq k\leq l$), where $\gamma_{th,k}$ represents the targeted threshold SINR of $U_k$, the OP at $U_l$ is formulated as
\begin{equation}\label{eq:7}
P_{out,l}=1-Pr\left(\left\lbrace \gamma_{SRU_{1\rightarrow l}}>\gamma_{th,1} \right\rbrace\cap\cdots\cap\left\lbrace \gamma_{SRU_{l\rightarrow l}}>\gamma_{th,l} \right\rbrace \right) 
\end{equation}
Then, with the help of (\ref{eq:5}), the event $\left\lbrace \gamma_{SRU_{k\rightarrow l}}>\gamma_{th,k} \right\rbrace $ can be expressed as
\begin{equation}\label{eq:8}
\begin{split}
&\left\lbrace \gamma_{SRU_{k\rightarrow l}}>\gamma_{th,k} \right\rbrace \\
&=\left\lbrace(B-2\vartheta_1\delta_k)A>\dfrac{2(\vartheta_2\bar{\gamma}B+\vartheta_3\bar{\gamma}C+\vartheta_4\bar{\gamma}^2BC+\vartheta_5)\delta_k}{\bar{\gamma}} \right\rbrace \\
&=\left\lbrace A>\dfrac{2(\vartheta_2\bar{\gamma}B+\vartheta_3\bar{\gamma}C+\vartheta_4\bar{\gamma}^2BC+\vartheta_5)\delta_k}{\bar{\gamma}(B-2\vartheta_1\delta_k)},B>2\vartheta_1\delta_k  \right\rbrace
\end{split} 
\end{equation}
which is obtained under the condition of $a_k-\gamma_{th,k}\sum_{t=k+1}^{L}a_t>0$ and the notation $\delta_k\stackrel{\triangle}{=}\frac{\gamma_{th,k}}{\bar{\gamma}(a_k-\gamma_{th,k}\sum_{t=k+1}^{L}a_t)}$ is used for mathematical tractability. If (\ref{eq:8}) is substituted into (\ref{eq:7}), the OP expression corresponding to $U_l$ can be written as
\begin{equation}\label{eq:9}
\begin{split}
&P_{out,l}=1- \\
&Pr\left(A>\dfrac{2(\vartheta_2\bar{\gamma}B+\vartheta_3\bar{\gamma}C+\vartheta_4\bar{\gamma}^2BC+\vartheta_5)\delta_l^{\dag}}{\bar{\gamma}(B-2\vartheta_1\delta_l^{\dag})},B>2\vartheta_1\delta_l^{\dag} \right) 
\end{split}
\end{equation}
which is obtained by defining $\delta_l^{\dag}=\underset{1\leq k\leq l}{\max}\left\lbrace \delta_k \right\rbrace $. By solving the probability problem given by (\ref{eq:9}), OP of $U_l$ can be mathematically expressed as 
\begin{equation}\label{eq:10}
P_{out,l}=1-\int\limits_{y=2\vartheta_1\delta_l^{\dag}}^{\infty}\int\limits_{z=0}^{\infty}\int\limits_{x=a}^{\infty}f_{A}(x)f_{C}(z)f_{B}^{(l)}(y)dxdzdy
\end{equation}
where $a=\frac{2(\vartheta_2\bar{\gamma}y+\vartheta_3\bar{\gamma}z+\vartheta_4\bar{\gamma}^2yz+\vartheta_5)\delta_l^{\dag}}{\bar{\gamma}(y-2\vartheta_1\delta_l^{\dag})}$. In (\ref{eq:10}), $f_X(\cdot)$ and $f_X^{(l)}(\cdot)$ are used to represent the probability density functions (PDFs) of a random variable $X$ and its $l$th order statistic, respectively. The following theorem provides the exact OP related to the $l$th user.

\textit{Theorem 1:} By solving (\ref{eq:10}), the exact OP for the $l$th user can be derived as
\begin{equation}\label{eq:11}
\begin{split}
&P_{out,l}=1-Q_l\widetilde{\sum}\binom{n_B-2}{r}\binom{L-l}{k}\binom{l+k-1}{p}\binom{n_1}{t_3}\binom{t_3}{k_2} \\
&\binom{m_{RU_l}n_R+k_1-1}{t_4}\frac{n_B(n_B-1)}{\Gamma(m_{SR})}\left( \frac{m_{SR}}{\hat{\Omega}_{SR}}\right)^{2m_{SR}}\frac{(-1)^{r+k+p}}{t!n_1!} \\
&\frac{2^{-t-n-m_{SR}+1}\Gamma(m_{SR}+n+t)}{\Gamma(m_{RU_l}n_R)\Gamma(m_{RR})}\left( \frac{m_{RU_l}}{\hat{\Omega}_{RU_l}}\right)^{m_{RU_l}n_R}\left(\frac{m_{RR}}{\Omega_{RR}}\right)^{m_{RR}}  \\
&\beta_n(r,m_{SR})\beta_{k_1}(p,m_{RU_l}n_R)\left( 2\vartheta_1\delta_l^{\dag}\right)^{m_{RU_l}n_R+k_1-t_4-1} \\
&e^{-2\vartheta_1\delta_l^{\dag}\frac{m_{RU_l}(1+p)}{\hat{\Omega}_{RU_l}}}\left(\frac{2\delta_l^{\dag}}{\bar{\gamma}}\right)^{n_1}\left(\vartheta_3\bar{\gamma}+2\vartheta_1\vartheta_4\bar{\gamma}^2\delta_l^{\dag} \right)^{\frac{n_1-t_3+t_4+1}{2}} \\
&\left(\vartheta_4\bar{\gamma}^2 \right)^{t_3}\left(\frac{\vartheta_2}{\vartheta_4\bar{\gamma}} \right)^{t_3-t_2}\kappa_{t_1t_2}\left(\frac{2\delta_l^{\dag}s_{t_1}\hat{\Omega}_{RU_l}}{\bar{\gamma}m_{RU_l}(1+p)}\right)^{\frac{t_3+t_4-n_1+1}{2}} \\
&e^{-2\delta_l^{\dag}s_{t_1}\vartheta_2}s_{t_1}^{n_1-t_2}\Phi_l,
\end{split} 
\end{equation} 
where 
\begin{equation}
\begin{split}
\widetilde{\sum}\equiv&\sum\limits_{r=0}^{n_B-2}\sum\limits_{n=0}^{r(m_{SR}-1)}\sum\limits_{t=0}^{m_{SR}-1}\sum\limits_{k=0}^{L-l}\sum\limits_{p=0}^{l+k-1}\sum\limits_{k_1=0}^{p(m_{RU_l}n_R-1)}\sum\limits_{t_4=0}^{m_{RU_l}n_R+k_1-1} \\ &\sum\limits_{t_1=1}^{T}\sum\limits_{t_2=1}^{\alpha_{t_1}}\sum\limits_{n_1=0}^{t_2-1}\sum\limits_{t_3=0}^{n_1}\sum\limits_{k_2=0}^{t_3}, \nonumber
\end{split} 
\end{equation} 
\begin{equation}
\begin{split}
&\Phi_l=\int\limits_{z=0}^{\infty}z^{k_2+m_{RR}-1}\left(\pi_l(z)\right)^{\frac{n_1-t_3+t_4+1}{2}}e^{-z\left(2\delta_l^{\dag}\vartheta_4\bar{\gamma}s_{t_1}+\frac{m_{RR}}{\Omega_{RR}} \right)} \\
&K_{t_3+t_4-n_1+1}\left(2\sqrt{\frac{2\delta_l^{\dag}m_{RU_l}(1+p)(\vartheta_3\bar{\gamma}+2\vartheta_1\vartheta_4\bar{\gamma}^2\delta_l^{\dag}s_{t_1})}{\bar{\gamma}\hat{\Omega}_{RU_l}}\pi_l(z)}\right)dz,   \nonumber
\end{split} 
\end{equation} 
In (\ref{eq:11}), $K_v(\cdot)$ denotes the $v$th-order modified Bessel function of the second kind \cite[eq.(8.407.1)]{Gradshteyn} and $\pi_l(z)=z+\frac{2\vartheta_1\vartheta_2\bar{\gamma}\delta_l^{\dag}+\vartheta_5}{\vartheta_3\bar{\gamma}+2\vartheta_1\vartheta_4\bar{\gamma}^2\delta_l^{\dag}}$. 

\textit{Proof: Please see Appendix A}.

\subsection{Lower Bound Analyses}
Unfortunately, there is no closed form solution of the integral $\Phi_l(z)$ in (\ref{eq:11}). However, an approximated form of (\ref{eq:11}) can be derived by upper-bounding the SINR given in (\ref{eq:5}). By applying some mathematical manipulations, (\ref{eq:9}) can be approximated as   
\begin{equation}\label{eq:12}
P_{out,l}\approx1-
Pr\left(\dfrac{\dfrac{\vartheta_1^{\prime}}{\vartheta_3^{\prime}} W\dfrac{\vartheta_2^{\prime}}{\vartheta_3^{\prime}}\bar{\gamma}B}{\dfrac{\vartheta_1^{\prime}}{\vartheta_3^{\prime}}W + \dfrac{\vartheta_2^{\prime}}{\vartheta_3^{\prime}}\bar{\gamma}B}>2\delta_l^{\dag}\bar{\gamma}\dfrac{\vartheta_1^{\prime}\vartheta_2^{\prime}}{\vartheta_3^{\prime}}\right) 
\end{equation}
where a new random variable $W\stackrel{\triangle}{=}\bar{\gamma}A/\left( \bar{\gamma}C+\vartheta_4^{\prime}\right) $ is defined for mathematical tractability. Also, $\vartheta_1^{\prime}=\dfrac{\bar{\gamma}}{2}\dfrac{\sigma_{RU_l}^2}{\rho_{RU_l}^2}+\dfrac{1}{\rho_{RU_l}^2}$, $\vartheta_2^{\prime}=\dfrac{1}{\rho_{SR}^2}$, $\vartheta_3^{\prime}=\dfrac{\bar{\gamma}\sigma_{RU_l}^2+1}{\rho_{SR}^2\rho_{RU_l}^2}$ and $\vartheta_4^{\prime}=\dfrac{\bar{\gamma}^2}{2}\sigma_{SR}^2+1$. Then, the left hand side of the argument of $Pr(\cdot)$ can be upper-bounded by using the inequality $xy/(x+y)\leq\min(x,y)$. The right hand side of this property also provides the lower-bound for the OP as
\begin{equation}\label{eq:13}
\begin{split}
P_{out,l}^{low}&=1-
Pr\left(\min\left(W\dfrac{\vartheta_1^{\prime}}{\vartheta_3^{\prime}},\bar{\gamma}B\dfrac{\vartheta_2^{\prime}}{\vartheta_3^{\prime}} \right)  >2\delta_l^{\dag}\bar{\gamma}\dfrac{\vartheta_1^{\prime}\vartheta_2^{\prime}}{\vartheta_3^{\prime}}\right) \\
&=1-\overline{F}_{W}\left(2\delta_l^{\dag}\bar{\gamma}\vartheta_2^{\prime} \right)\overline{F}_{B}^{(l)}\left(2\delta_l^{\dag}\vartheta_1^{\prime} \right)  
\end{split}
\end{equation}
where $F_X(\cdot)$ and $\overline{F}_X(\cdot)$ represent the cumulative distribution function (CDF) of a random variable $X$ and its complementary, respectively. Also, $F_X^{(l)}(\cdot)$ denotes the $l$th order statistic of $X$. Finally, if the complementaries of CDF expressions provided in the following theorem are substituted into (\ref{eq:13}), lower-bound of the OP for $l$th user can be obtained in closed-form.

\textit{Theorem 2:} The CDF expressions of random variables predefined as $W$ and $B$ which also denote SINR and SNR of the first and second hops can be derived respectively as
\begin{equation}\label{eq:14}
\begin{split}
&F_{W}(x)=1-\widetilde{\widetilde{\sum}}\binom{n_B-2}{r}\binom{t_4}{t_5}\frac{n_B(n_B-1)}{\Gamma(m_{SR})\Gamma(m_{RR})}\left( \dfrac{m_{SR}}{\hat{\Omega}_{SR}}\right)^{2m_{SR}} \\
&\left(\dfrac{m_{RR}}{\Omega_{RR}}\right)^{m_{RR}}\dfrac{(-1)^{r}}{t!t_4!}\frac{\beta_n(r,m_{SR})\Gamma(m_{SR}+n+t)}{2^{t+n+m_{SR}}}\Gamma(t_5+m_{RR}) \\
&\left(\dfrac{\vartheta_4^{\prime}}{\bar{\gamma}}\right)^{t_4-t_5}\kappa_{t_1t_2}s_{t_1}^{t_4-t_2}x^{t_4}e^{-s_{t_1}x\frac{\vartheta_4^{\prime}}{\bar{\gamma}}}\left(s_{t_1}x+\frac{m_{RR}}{\Omega_{RR}} \right)^{-t_5-m_{RR}},
\end{split} 
\end{equation}
\begin{equation}\label{eq:15}
\begin{split}
F_{B}^{(l)}(x)&=1-Q_l\widetilde{\widetilde{\widetilde{\sum}}}\binom{L-l}{k}\binom{l+k}{p}\dfrac{(-1)^{k+p-1}}{l+k} \\
&\beta_{k_1}(p,m_{RU_l}n_R)x^{k_1}e^{-x\frac{m_{RU_l}}{\hat{\Omega}_{RU_l}}}, \\
\end{split} 
\end{equation}
where $\widetilde{\widetilde{\sum}}\equiv\sum_{r=0}^{n_B-2}\sum_{n=0}^{r(m_{SR}-1)}\sum_{t=0}^{m_{SR}-1} \sum_{t_1=1}^{T}\sum_{t_2=1}^{\alpha_{t_1}}\sum_{t_4=0}^{t_2-1}\sum_{t_5=0}^{t_4}$ and $\widetilde{\widetilde{\widetilde{\sum}}}\equiv\sum_{k=0}^{L-l}\sum_{p=1}^{l+k}\sum_{k_1=0}^{p(m_{RU_l}n_R-1)}$.

\setcounter{equation}{19}
\begin{figure*}[!t]
	\begin{equation}\label{eq:21}
		\begin{array}{c}
			\Xi_1=\left[ \sum\limits_{t=0}^{m_{SR}-1}\dfrac{n_B(n_B-1)\Gamma(m_{SR}(n_B-1)+t)\Gamma(m_{SR}n_B+m_{RR})}{\Gamma(m_{SR})(\Gamma(m_{SR}+1))^{n_B-2}t!2^{-m_{SR}+t}\Gamma(m_{SR}n_B+1)\Gamma(m_{RR})}\left(\dfrac{m_{SR}\varLambda_l^{\dag}}{m_{RR}\Omega_{SR}} \right)^{m_{SR}n_B} \right]^{-\frac{1}{(1-\mu)m_{SR}n_B}} \\
			\Xi_2=\left[\binom{L}{l}\frac{1}{(\Gamma(m_{RU_l}n_R+1))^l} \right]^{-\frac{1}{m_{RU_l}n_R l}}\frac{\Omega_{RU_l}}{2\varLambda_l^{\dag}m_{RU_l}}
		\end{array}
	\end{equation}
	\noindent\rule{\textwidth}{.5pt}
\end{figure*}
\setcounter{equation}{15} 

On the other hand, in case of ideal conditions, $\vartheta_1^{\prime}=\vartheta_2^{\prime}=\vartheta_3^{\prime}=\vartheta_4^{\prime}=1$ and the random variable $W$ can be approximated as $W\simeq A/C$. Therefore, $F_{W}(x)$ can be obtained for the ideal case as
\begin{equation}\label{eq:16}
\begin{split}
&F_{W}(x)=1-\widetilde{\widetilde{\sum}}\binom{n_B-2}{r}\frac{n_B(n_B-1)}{\Gamma(m_{SR})\Gamma(m_{RR})}\left( \dfrac{m_{SR}}{\Omega_{SR}}\right)^{2m_{SR}} \\
&\left(\dfrac{m_{RR}}{\Omega_{RR}}\right)^{m_{RR}}\dfrac{(-1)^{r}}{t!t_4!}\frac{\beta_n(r,m_{SR})\Gamma(m_{SR}+n+t)}{2^{t+n+m_{SR}}} \\
&\Gamma(t_4+m_{RR})\kappa_{t_1t_2}s_{t_1}^{t_4-t_2}x^{t_4}\left(s_{t_1}x+\frac{m_{RR}}{\Omega_{RR}} \right)^{-t_4-m_{RR}},
\end{split} 
\end{equation} 
where $\widetilde{\widetilde{\sum}}\equiv\sum\limits_{r=0}^{n_B-2}\sum\limits_{n=0}^{r(m_{SR}-1)}\sum\limits_{t=0}^{m_{SR}-1} \sum\limits_{t_1=1}^{T}\sum\limits_{t_2=1}^{\alpha_{t_1}}\sum\limits_{t_4=0}^{t_2-1}$. Also, the CDF of random variable $B$ remains the same for ideal case.  

\textit{Proof: Please see Appendix B}.

\subsection{Asymptotic Analyses}
In this subsection, in order to provide more meaningful insights into the system performance, we investigate the OP in high SNR region by deriving simple theoretical expressions including diversity order and array gain metrics. The analyses are carried out in the following two subsections.

\subsubsection{In Case of Ideal Conditions}
In the absence of CEE and FBD effects, in high SNR region (when $\bar{\gamma}\rightarrow\infty$), the investigated system enjoys diversity order and array gain in case that the quality of SI cancellation satisfies $\mu\neq 1$. Therefore, (\ref{eq:9}) can be asymptotically expressed as
\begin{equation}\label{eq:17}
\begin{split}
P_{out,l}^{\infty}&=\underset{\bar{\gamma}\rightarrow\infty}{\lim}Pr\left(\dfrac{\bar{\gamma}AB}{\bar{\gamma}A+\left(\bar{\gamma}B+1 \right)\left(\bar{\gamma}C+1 \right)  }\leq 2\delta_l^{\dag} \right) \\
&\approx\underset{\bar{\gamma}\rightarrow\infty}{\lim}Pr\left(\dfrac{W\bar{\gamma}B}{W+\bar{\gamma}B}\leq 2\varLambda_l^{\dag} \right) \\
&\approx\underset{\bar{\gamma}\rightarrow\infty}{\lim}Pr\left(\min\left(W,\bar{\gamma}B \right)\leq 2\varLambda_l^{\dag}  \right) \\
&=F_{W}^{\infty}\left( 2\varLambda_l^{\dag}\right) +F_{B}^{(l),\infty}\left(\frac{2}{\bar{\gamma}}\varLambda_l^{\dag}\right), 
\end{split}
\end{equation}
where $W\simeq A/C$ (since the analysis is conducted for the ideal case) and $\varLambda_l^{\dag}\stackrel{\triangle}{=}\frac{\gamma_{th,k}}{a_k-\gamma_{th,k}\sum_{t=k+1}^{L}a_t}$ are defined. Then, with the help of high SNR approximation approach proposed in \cite{ZWang}, asymptotic OP in (\ref{eq:17}) can be written as $P_{out,l}^{\infty}\approx\left(G_{ag}\bar{\gamma} \right)^{-G_{do}}+\textit{O}\left( \bar{\gamma}^{-G_{do}}\right)$, where $G_{do}$ and $G_{ag}$ represent diversity order and array gain metrics, respectively, and $\textit{O}(\cdot)$ denotes negligible high order terms. 

\textit{Theorem 3:} If the asymptotic CDFs $F_{W}^{\infty}\left(x\right)$ and $F_{B}^{(l),\infty}\left(x\right)$ are derived and substituted into (\ref{eq:17}), the asymptotic OP of the $l$th user can be obtained such that diversity order and array gain metrics are respectively given as
\begin{equation}\label{eq:19}
G_{do}=\min\left\lbrace \left(1-\mu\right)m_{SR}n_B,m_{RU_l}n_R l\right\rbrace,  
\end{equation}
\begin{equation}\label{eq:20}
G_{ag}=\left\{
\begin{array}{ll}
\Xi_1 & (1-\mu)m_{SR}n_B<m_{RU_l}n_R l \\
\Xi_2 & (1-\mu)m_{SR}n_B>m_{RU_l}n_R l \\
\Xi_1+\Xi_2 & (1-\mu)m_{SR}n_B=m_{RU_l}n_R l \\
\end{array}
\right.,
\end{equation} 
where $\Xi_1$ and $\Xi_2$ are given at top of this page. 

\textit{Proof: Please see Appendix C}.

When the quality of SI cancellation satisfies $\mu=1$, the effect of SI channel becomes much more dominant in the system SINR compared to the $R-U_l$ link. Thus, the effect of $R-U_l$ link can be neglected in the high SNR region. Therefore, asymptotic OP can be expressed as $P_{out,l}^{\infty}\approx1-Pr(W>2\delta_l^{\dag}\bar{\gamma})=F_{W}^{\infty}( 2\varLambda_l^{\dag})$ which is independent from the average SNR and causes an error floor even without any CEE and FBD effects.    

\subsubsection{In Case of Practical Conditions}
In the presence of CEE and FBD effects, in the high SNR region, the predefined constant values of $\vartheta$ given by (\ref{eq:6}) can be approximated as $\vartheta_1\approx\frac{\bar{\gamma}\sigma_{RU_l}^2}{2\rho_{RU_l}^2}$, $\vartheta_2\approx\frac{\bar{\gamma}\sigma_{SR}^2}{2\rho_{SR}^2}$, $\vartheta_3\approx\frac{\bar{\gamma}\sigma_{RU_l}^2}{\rho_{SR}^2\rho_{RU_l}^2}$, $\vartheta_4=\frac{1}{\rho_{SR}^2}$, $\vartheta_5\approx\frac{\bar{\gamma}^2\sigma_{SR}^2\sigma_{RU_l}^2}{2\rho_{SR}^2\rho_{RU_l}^2}$. Then, by substituting these values into (\ref{eq:11}), asymptotic OP can be calculated for practical conditions in all values of $\mu$. Eventually, the system suffers from an error floor, therefore diversity order analysis can not be conducted as in the ideal case.      

\section{TEST-BED IMPLEMENTATION}

\subsection{Hardware Components and Software Structure}

\begin{figure*}[!t]
	\centering
	\includegraphics[scale=0.08]{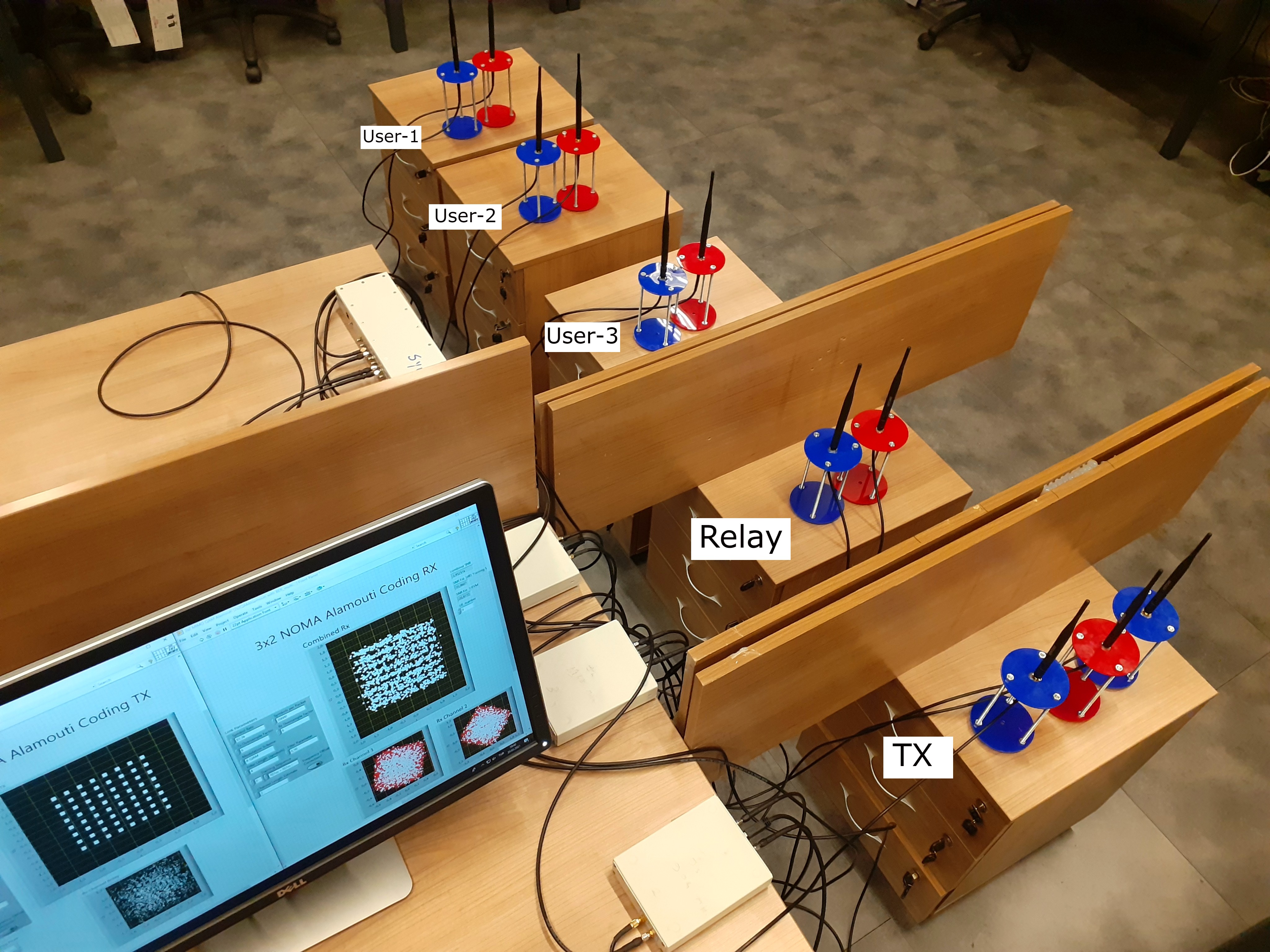}
	\setlength{\belowcaptionskip}{-2pt}
	\caption{Test-bed set-up model.}
	\label{fig:1_1}       
\end{figure*}

In order to support the theoretical study, 3-user dual-hop Alamouti coded NOMA tests are carried out using USRP SDRs. While a USRP-2901 (two RF channels) and a USRP-2900 (single RF channel) are used to represent the BS with 3 antennas, a USRP-2901 is used for both users, and a relay with 2 antennas as shown in Fig. \ref{fig:1_1}\footnote{Test-bed implementation was realized at \.{I}stanbul Technical University Wireless Communications Research Laboratory.}. Each radio has an operating frequency range of 70 MHz to 6 GHz, the maximum output power of 20 dBm in the transmitter, and a maximum input power of -15 dBm in the receiver. It should be noted that all devices have noise figures up to 5 to 7 dB. Antennas connected to these radios are placed at appropriate distances using SMA extension cables, which also act as a potential noise source. In order to minimize the exposure from the external environment, the connections of the devices with the host are provided via universal serial bus (USB) 3.0 instead of an Ethernet switch. In addition to the SDRs, a CDA-2990 Octoclock device with a global positioning system (GPS) disciplined oscillator is used to regulate the timing and frequency synchronization of all radios of the system. A computer with an Intel-Xeon E5-1620 processor is used for programming the radios, recording, and displaying the data obtained during tests. With LabVIEW graphical modeling tool, the diagram flow of the analytical study is converted into the algorithm, so tests that will transfer data in real-time have been designed. As it can be seen in the test environment figure, the radio antennas are located such that the distance from  BS to the relay is 35 cm, distance between the closest user and relay is 52 cm and the distance among users is 52 cm. Two pieces of 1m x 4m wood panels covered with silicon are placed between BS-Relay and Relay-Users with the purpose of having Rayleigh fading for all channels by obstructing the line of sight.

\subsection{Test Configuration}

Dual-hop 3x2 Alamouti tests consist of 3 stages and the operating frequency and bandwidth for all stages are 2.11 GHz and  400kHz respectively. Tests are performed with a single carrier and 4-QAM modulation. Receivers and transmitters with an oversampling factor of 8 receive 500,000 IQ data every second (total 1M), which makes the number of symbols sent every second is 62,500. With each packet decided to contain 20,000 symbols, it takes approximately 0.32 seconds to transmit each one of them. Also, power coefficients are selected as 0.761 (User-1), 0.191, and 0.048 (User-3). A summary of the parameters of the system is given in Table \ref{Table1}.

To briefly describe the test steps; In the first stage, the best 2 of 3 antennas are selected by applying the transmitter antenna selection method. In the second stage, 3-user NOMA symbols are encoded with the Alamouti scheme and transmitted to the relay. In the third stage, the sampled data recorded in the relay is transferred to the users without any additional processing. Users perform outage analysis by extracting their own data after applying SIC to the data they obtain. This data transfer scheme occurs in large numbers but in short periods of time, as shifts and weakening of the sequential time columns in the Alamouti matrix will cause large-scale errors in the decoding process. In-depth details of the stages are explained in the following paragraph. 

As the first stage of the tests, 2 of 3 antennas with the highest channel gain transmitter antennas are selected as they perform transmitter antenna selection tests, hereby; one transmitter antenna is eliminated. It should be noted that the antenna selection process has been made in the Rayleigh fading channel setup and only for transmitting antennas.

\begin{table}[!t]
	\caption{Test-bed environment configurations.}
	\centering
	\begin{tabular}{|c|c|c|}
		\hline
		\textbf{Parameters}& \multicolumn{2}{l|}{\textbf{Values}} \\
		\hline
		Modulation         & \multicolumn{2}{l|}{4-QAM} \\
		\hline
		Carrier Frequency  & \multicolumn{2}{l|}{2.11 GHz} \\
		\hline
		Bandwidth          & \multicolumn{2}{l|}{400 kHz} \\
		\hline
		I/Q Data Rate      & \multicolumn{2}{l|}{500 kS/sec} \\
		\hline
		Channel Fading     & \multicolumn{2}{l|}{Nakagami-$m$, $m=0.98$} \\
		\hline
		Power Coefficients & \multicolumn{2}{l|}{$0.761$, $0.191$, $0.048$} \\
		\hline
		Transmitter Gain   & \multicolumn{2}{l|}{20 to 40 dB} \\
		\hline
		Receiver Gain      & \multicolumn{2}{l|}{15 dB} \\
		\hline
		Threshold Values (far to near)  & \multicolumn{2}{l|}{$2.0$, $2.5$, $3.0$} \\
		\hline
		\multirow{4}{*}{Distance From Relay to} & BS     & 35 cm \\ \cline{2-3} 
		& User 3 & 52 cm \\ \cline{2-3}
		& User 2 & 104 cm \\ \cline{2-3}
		& User 1 & 156 cm \\
		\hline
		\multirow{4}{*}{Number of Antennas}     & BS     & 3 \\ \cline{2-3} 
		& Relay  & 2 \\ \cline{2-3}
		& Users  & 2 \\
		\hline
	\end{tabular}
	\label{Table1}
\end{table}

In the second stage, first hop tests are performed and the sampled data in the relay are saved. Using random seed for three users in the base station, bits of 3 users are produced via a pseudo-random number generator before 4-QAM modulation is applied separately for each user. Then; the resulting symbol streams are multiplied by the roots of the power coefficient of each user, the sum of which results in a superposition symbol. By applying the Alamouti STBC algorithm to this symbol, 2-dimensional symbols with 2 baseband streams (each for an antenna) are created. Then, the repetitive training symbol sequence is prepended to each stream for synchronization operations at the receiver. Finally, the data is upsampled to be sent, creating a pulse train and a pulse shaping filter is applied. Thus, the data leaves the BS and reaches over the relay which has a single operation such that saving the incoming data and transmitting the saved data. The relay receiver gain is kept constant at 35 dB and the transmitter gain is set to 40 dB during tests and its impact is captured with SNR. On the second hop, the data coming out of the relay reaches users located at three different distances. For these stationary users, the weak user is assumed as the furthest one, while the strong user is assumed to be the closest one to the relay. Users begin to recover data where each row of the 2-dimensional array they receive represents the stream of each antenna. The first thing to do after the symbol passes through the matched filter is to compare the energy of the streams of each antenna and decide on the stronger antenna. In this way, the symbol timing offset correction operations to be done immediately afterward are done by using the training symbol sequence belonging to the strong stream. Substantially, the head of the symbol can be found at the point where the correlation in the training symbol belonging to a single stream is maximum. 

Frame synchronization procedures of Schmidl and Cox are applied in order to synchronize both antennas after these processes. Similar to the symbol synchronization process aforementioned, only, this particular synchronization process can be completed once the correlation of both streams of the two antennas is at the maximum.  It should be noted that fine frequency correction procedures are not applied to the system that resolves the coarse frequency offset using Octoclock. When it comes to the channel estimation and equalization, it is assumed that there are 4 channels in total for the 2x2 Alamouti setup and 4 channel estimation operations are performed for each hypothetical channel. With 4 channel vectors, coefficients are obtained by using linear LS estimators for 4 training symbols and, in parallel with this purpose, training symbols are extracted from streams. At this point, linear SNR value is obtained by the products of the sum of the channel coefficient matrix and the transmission power ($P_t\cdot|h_{ij}|^2$) at the same time. The symbol matrix, where only consists of unequalized data, is converted into a 1-size vector with Alamouti decode scheme. By using the previously obtained 4 channel estimation coefficients, the MRC process is applied 2 times and then the data of the first user is obtained (Test-bed photo, Fig. \ref{fig:1_1}, and interface belong to the first user test). The first user can simply jump into outage performance analysis after power normalization, while the second and third users need to apply SIC operations in addition to the first user's operations to get their own data. So, the second user implements single-stage SIC operations, while the strong third user implements 2-layer SIC operations. After users obtain their data, as a conventional outage analysis, the SNR values of the users are compared with the predefined threshold values and the packets below the threshold are marked as an outage. As a result, the ratio of the outage cases to the SNR vector gives the OP value.

Synchronization operations are deadly crucial in data recovery processes. It is fiendishly difficult to synchronize the system without using a common external time and frequency reference. During the test taken indoors, the problems that can create hardware impairment depending on external factors (difference in antenna height or SMA cable length, inequality of distance between common antennas etc.) have been solved as much as possible, and the parameters of the system have been selected by considering these effects. In order to prevent data loss and to avoid buffer overflows, tests of each user are taken one by one. In addition to all these, the channel distribution estimation test of the test environment evaluated to be 0.98 of shape parameter with the maximum likelihood estimation of Nakagami distribution by using the channel coefficients.

\section{Numerical Results}
In this section, numerical results corresponding to the investigated system verified by Monte Carlo simulations and SDR tests are illustrated. The system is considered as consisting of three mobile users ($L=3$) in order to be an example. Unless otherwise stated, in all figures, $SNR=\bar{\gamma}=P/\sigma^2$, $\eta=4$, $\alpha=1$ as in \cite{Duarte,Rodriguez}, power allocation coefficients and target SINRs  related to mobile users are assumed as $a_1=1/2$, $a_2=1/3$, $a_3=1/6$, and $\gamma_{th,1}=0.9$, $\gamma_{th,2}=1.5$ and $\gamma_{th,3}=2$, respectively. Also, markers illustrate simulation results, normalized distance between the BS and relay is fixed as $d_{SR}=0.5$ while between the relay and mobile users as $d_{RU_1}=d_{RU_2}=d_{RU_3}=0.5$. $m_{SR}$, $m_{RR}$ and $m_{RU}$ denote Nakagami-$m$ parameter of $S-R$, $R-R$ and $R-U_l$ links ($m_{RU_1}=m_{RU_2}=m_{RU_3}=m_{RU}$ is assumed for simplicity). In order to fairly compare the investigated FD-NOMA system with FD-OMA and HD-NOMA counterparts, target SNRs can be obtained by following relationships $\log_2(1+\gamma_{th})=\sum_{i=1}^{L}\log_2(1+\gamma_{th,j})$ and $\frac{1}{2}\log_2(1+\gamma^{HD}_{th,l})=\log_2(1+\gamma^{FD}_{th,l})$, respectively. Lower bound (LB), asymptotic (Asymp) and theoretical (Theo) abbreviations are made for easy of reading. 

\begin{figure}[!b]
	\centering
	\includegraphics[width=0.85\columnwidth]{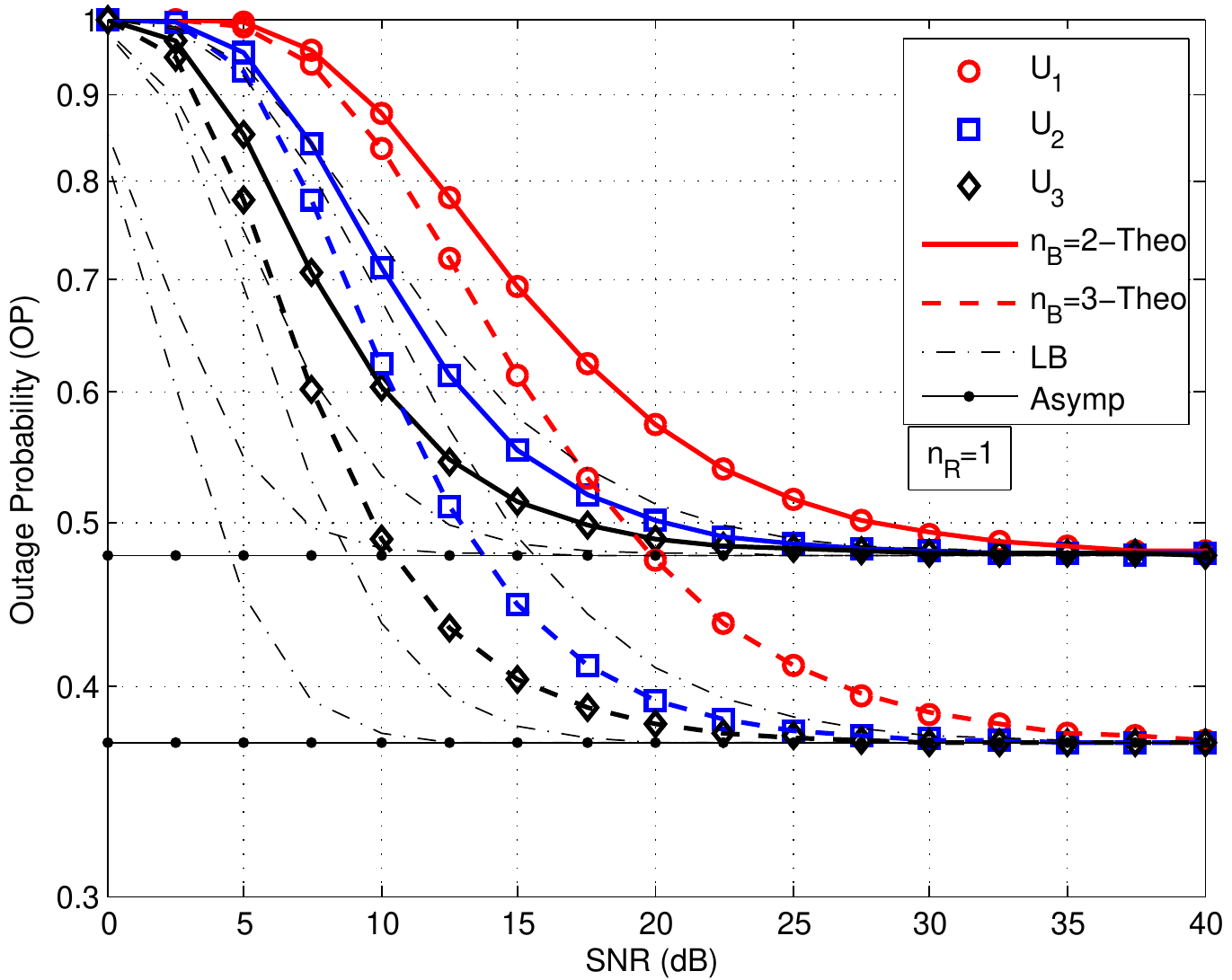}
	\caption{OP of the investigated system in case $\mu=1$ and $m_{SR}=m_{RR}=m_{RU}=1$ for ideal conditions.}
	\label{fig:2}
	\vspace{-2 mm}
\end{figure}  

Fig. 3 illustrates the OP performance of the investigated (TAS/Alamouti-MRC FD-NOMA system) in case that the value related to quality of SI cancellation in FD relay is set as $\mu=1$ (which means the worst scenario), $m_{SR}=m_{RR}=m_{RU}=1$ under ideal conditions (without any CEE and FBD effects). As clearly seen in the figure, the system suffers from an error floor (which indicates zero diversity) in high SNR region in the worst case of SI cancellation, even under ideal conditions, and OP curves of all users reach the same level. Also, lower bound curves are very close to exact results, even matches with them at high SNR values. However, level of the error floor can be decreased by increasing the number of transmit antennas at the BS, so performance of the system is significantly improved. 

In Fig. 4, OP curves of the investigated system are depicted for different values of $\mu$ and $m_{SR}=m_{RR}=m_{RU}=1$ in case of ideal conditions. We observe that OP performance of all users can be improved as the quality of SI cancellation gets better (as the value of $\mu$ decreases). Particularly, according to the first user, the difference of OP between values of $\mu=0.5$ and $\mu=0$ is slightly and OP performances are completely the same at high SNR values. However, the value of $\mu$ is mostly effective in the performance improvement of the second and third users, even in entire SNR region. On the other hand, the third user receives the best performance improvement thanks to the quality of SI cancellation process. When the value of $\mu$ is kept in the range of $0\leq\mu<1$, OP of all users exhibits diversity order and array gain which are directly effected by $\mu$ as well. These results are also verified by asymptotic analysis and can be observed from asymptotic curves which converge to slopes of exact results in the figure.

\begin{figure}[!t]
	\centering
	\includegraphics[width=0.85\columnwidth]{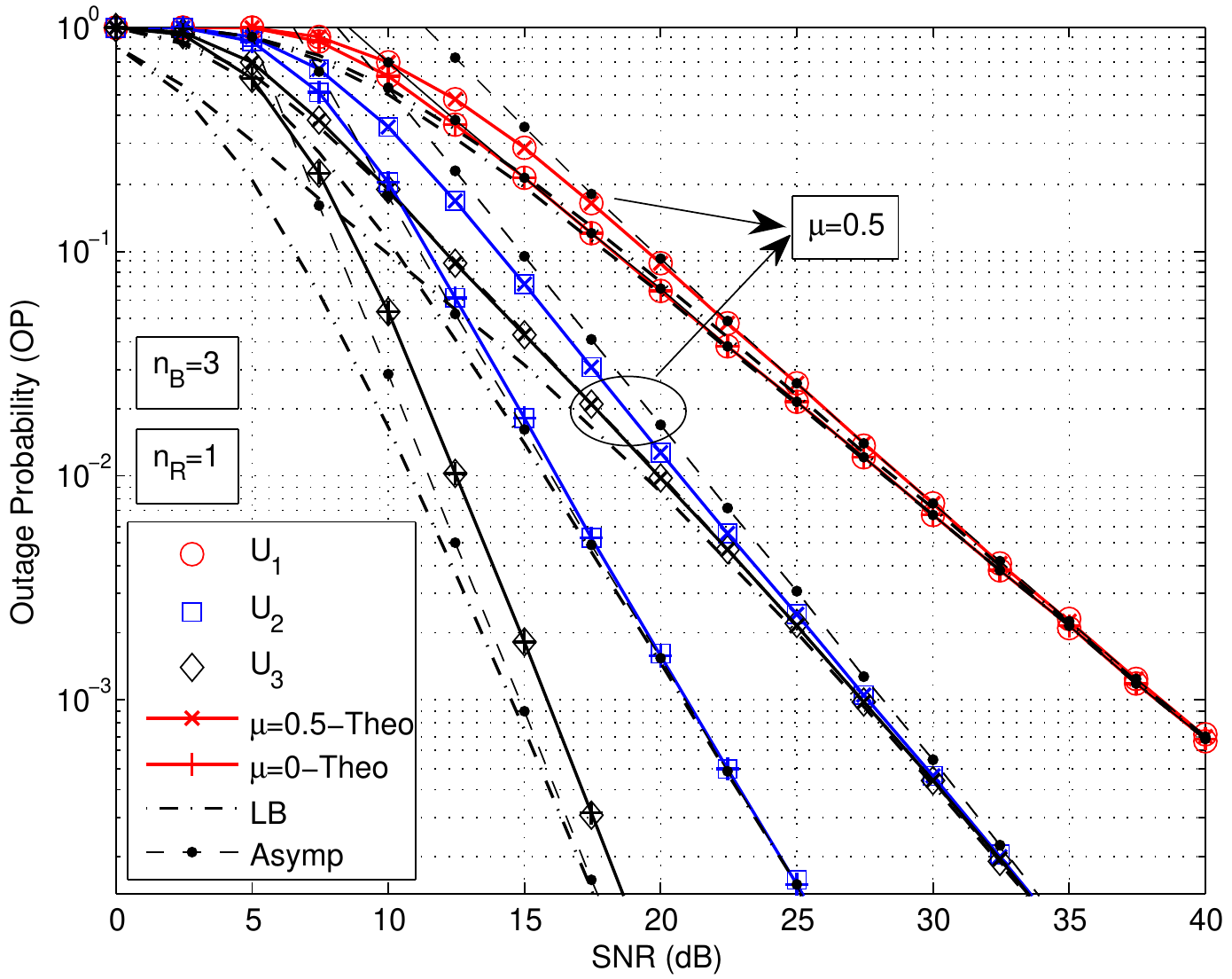}
	\caption{OP of the investigated system in case different values of $\mu$ and $m_{SR}=m_{RR}=m_{RU}=1$ for ideal conditions.}
	\label{fig:3}
	\vspace{-2 mm}
\end{figure}
\begin{figure}[!b]
	\centering
	\includegraphics[width=0.85\columnwidth]{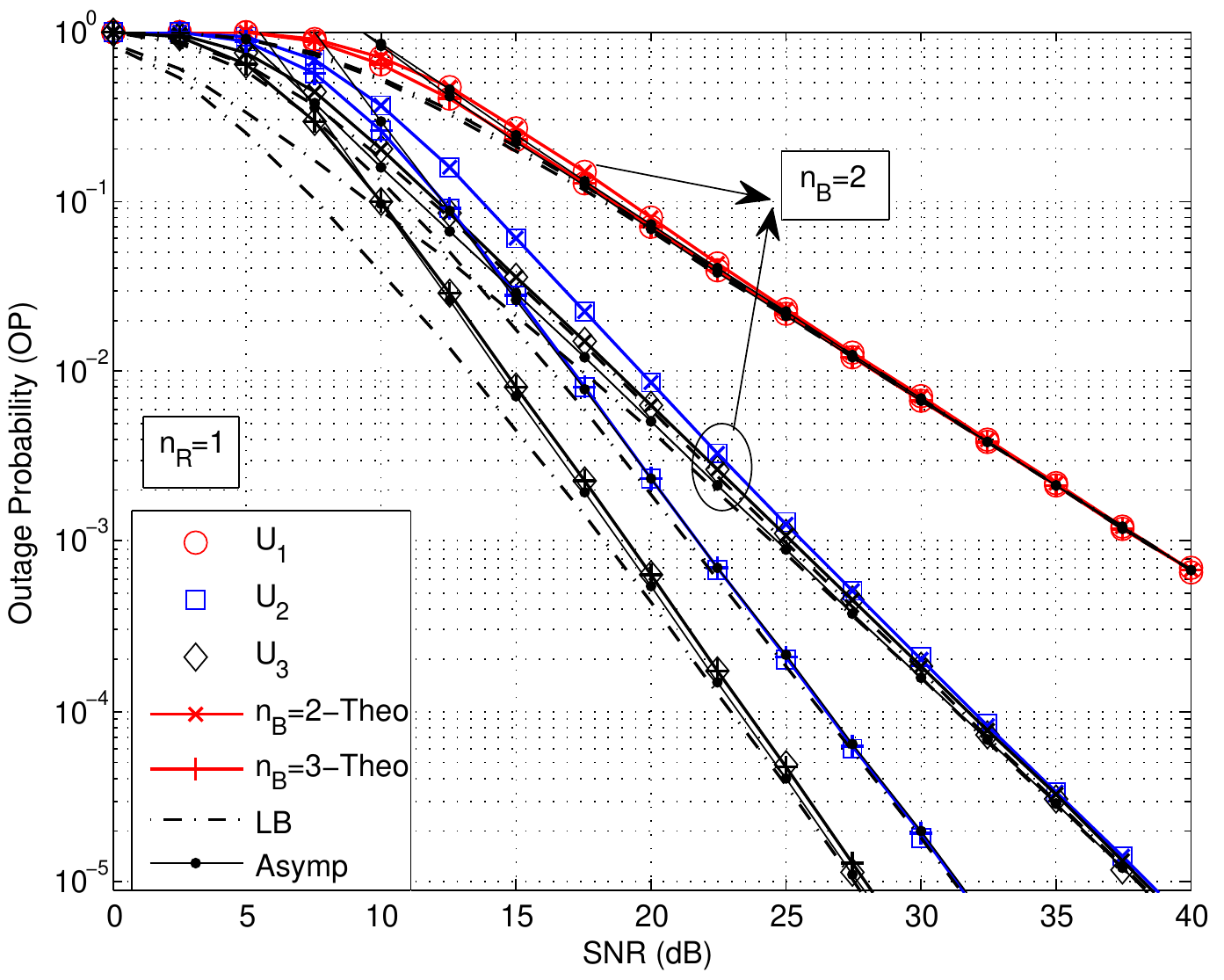}
	\caption{OP of the investigated system in case different values of $n_B$, $\mu=0.25$ and $m_{SR}=m_{RR}=m_{RU}=1$ for ideal conditions.}
	\label{fig:4}
	\vspace{-2 mm}
\end{figure}

Figs. 5 and 6 depict OP performances of the investigated system for ideal and practical conditions ($\sigma_{est,SR}^2=0.01$, $\sigma_{est,l}^2=0.01$ and $f_D\tau=0.03$), respectively, in case of $\mu=0.25$, $m_{SR}=m_{RR}=m_{RU}=1$ and different number of antennas at the BS and users. As can be clearly seen in Fig. 5, with the TAS scheme, OP performance of all users is significantly improved when the curves corresponding to results of $n_B=2$ and $n_B=3$ configurations are compared. Especially, for an OP value of $10^{-4}$, $6$ dB and $8$ dB SNR gains are obtained for the second and third users. On the other hand, from Fig. 6, error floors are observed due to the CEE and FBD effects, and also asymptotic curves obtained by theoretical analysis verify these results. Therefore, the system can not utilize diversity order and array gain in case of practical conditions. However, OP performance significantly gets better as the number of antennas at the BS and users increases. In addition, according to the first user, the TAS scheme has only effect on low SNR region in case of ideal conditions whereas it is quite effective at all SNR values under CEE and FBD effects.

\begin{figure}[!t]
	\centering
	\includegraphics[width=0.85\columnwidth]{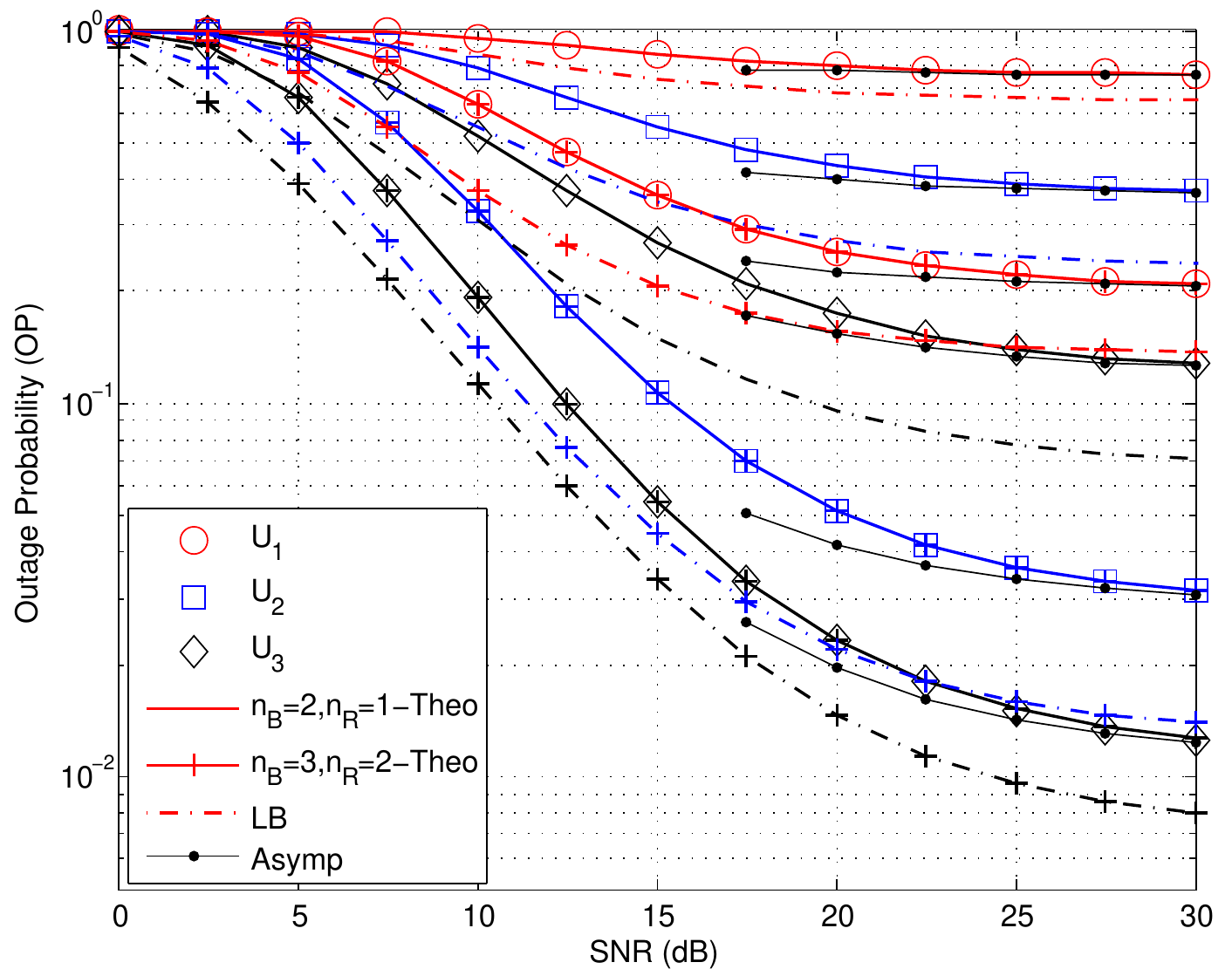}
	\caption{OP of the investigated system in case different values of $n_B$, $n_R$, $\mu=0.25$ and $m_{SR}=m_{RR}=m_{RU}=1$ in the presence of CEE and FBD.}
	\label{fig:5}
	\vspace{-2 mm}
\end{figure}
\begin{figure}[!t]
	\centering
	\includegraphics[width=0.85\columnwidth]{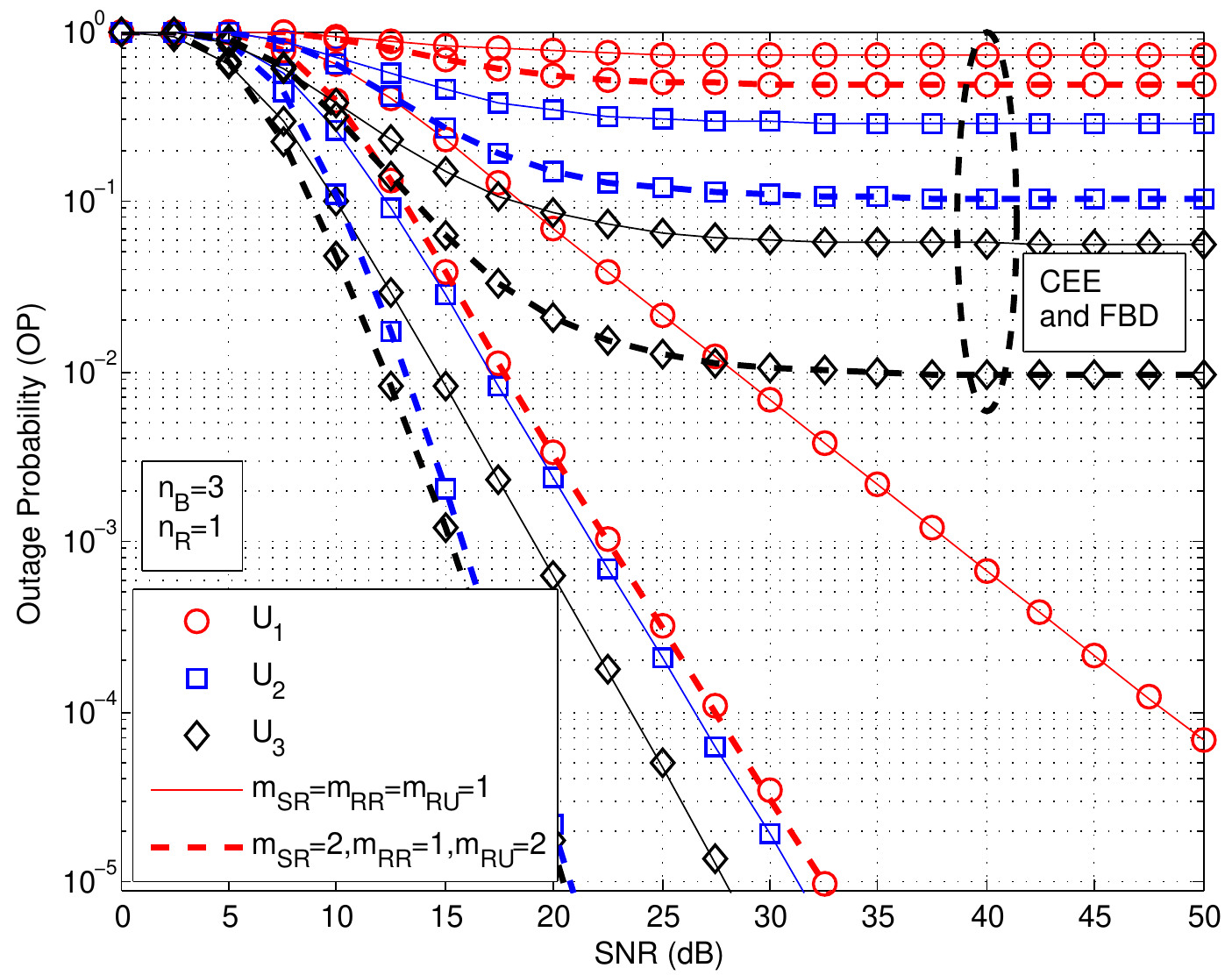}
	\caption{OP comparison of ideal and practical conditions of the investigated system for different channel parameters in case $\mu=0.25$.}
	\label{fig:6}
	\vspace{-2 mm}
\end{figure}

Fig. 7 shows OP performance comparisons of the investigated system for ideal and practical conditions in case of different Nakagami-$m$ channel parameters and $\mu=0.25$. For practical conditions, CEE and FBD parameters are set as $\sigma_{est,SR}^2=0.01$, $\sigma_{est,l}^2=0.01$ and $f_D\tau=0.03$. From the figure, it can be clearly observed that CEE and FBD effects seriously deteriorate the performance of the system. In addition, as the channel condition gets better (means the increasing of Nakagami-$m$ parameter), OP performances of all users increase. More specifically, for the determined configurations, the first and third users receive the highest and lowest performance gain, respectively, in ideal conditions whereas vice versa in practical manner. The main reason for the emergence of this result is the effect of diversity and array gains in ideal conditions.

Fig. 8 illustrates test-bed implementation results in terms of OP obtained through USRP SDRs to support the derived theoretical analyses and demonstrate feasibility of the system in real-life manner. Test-bed OP curves are obtained according to parameters in Table \ref{Table1} provided in Sect. IV. Besides, variances of CEEs in both hops are set as $\sigma_{est,SR}^2=\sigma_{est,l}^2=0.048$ while $f_D\tau=0$ (no FBD). Also, we assume that SI cancellation at FD relay can be performed at a good level and thus the corresponding parameter is $\mu=0$. We observe that the first user enjoys the best OP performance since the allocated power coefficient is quite high compared to the others. On the other hand, test-bed results exhibit the similar behavior with simulation and theoretical results. Also, test-bed curves are quite close to those of simulation and theoretical results in terms of all users up to certain SNR values, then diverge and error floor, frequently encountered in real-time tests occurs as a result of estimation process. This floor is seen due to the composite effect of the real-time design related to impairments. This result implies that the investigated system can be realized in a real-life manner by also validating the theoretical analyses.    

\begin{figure}[!t]
	\centering
	\includegraphics[width=0.85\columnwidth]{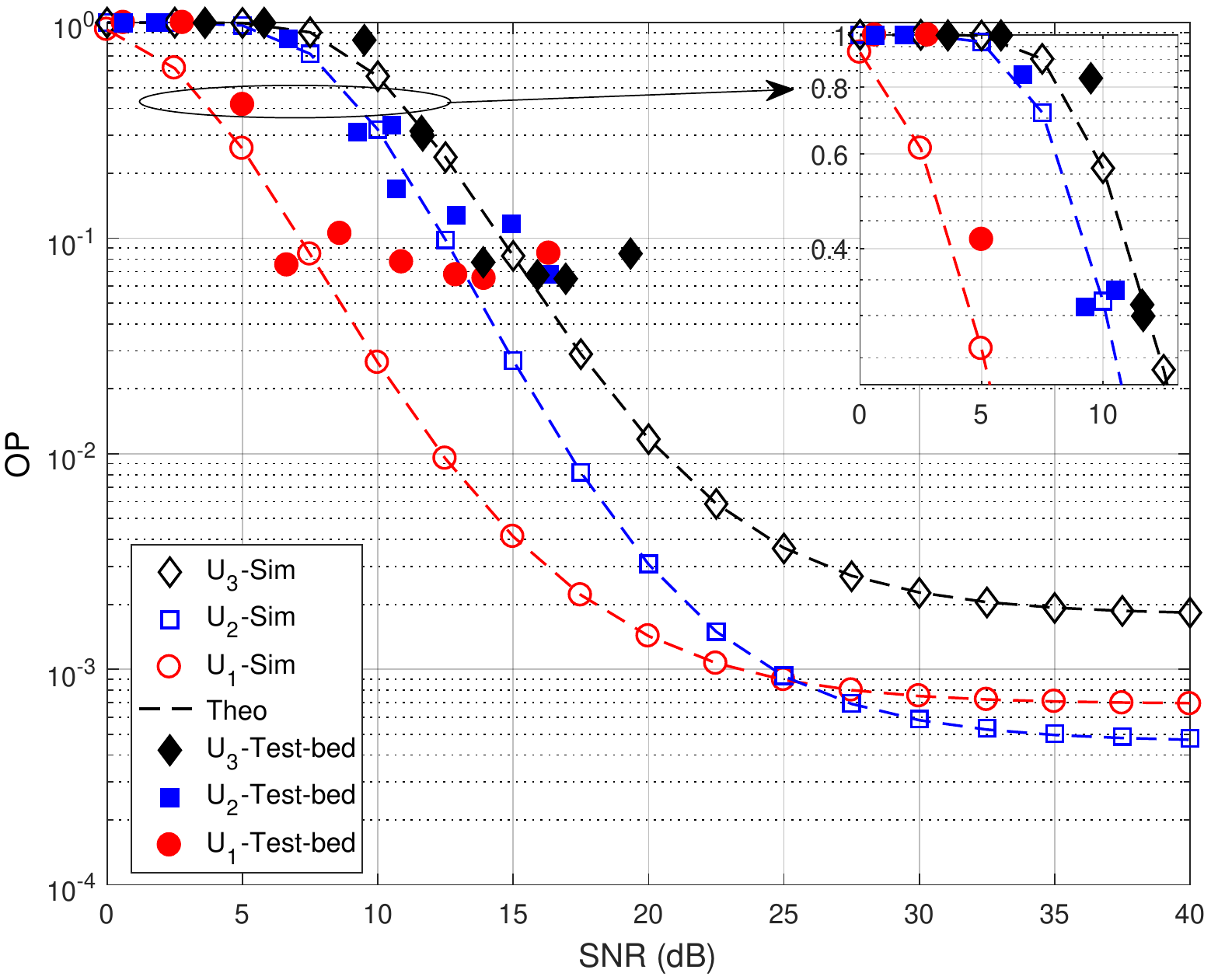}
	\caption{OP of the investigated FD-NOMA system with USRP SDR in a practical manner.}
	\label{fig:61}
	\vspace{-3 mm}
\end{figure}

Fig. 9 illustrates OP performance comparisons of the investigated FD-NOMA and HD-NOMA systems versus $\mu$ for fixed $SNR=15$ dB, $n_R=1$ and $m_{SR}=m_{RR}=m_{RU}=1$ in case ideal conditions. HD-NOMA curves are obtained through simulations. In order to have fair comparisons and satisfy the condition of $a_k>\gamma_{th,k}\sum_{t=k+1}^{L}a_t$ predefined in the exact OP analysis, we set threshold SNRs of HD-NOMA as $\gamma^{HD}_{th,1}=0.9$, $\gamma^{HD}_{th,2}=1.5$ and $\gamma^{HD}_{th,3}=2$. As observed from the figure, FD-NOMA outperforms HD-NOMA counterpart at all values of $\mu$ according to the first user. Besides, FD-NOMA provides better performance when the value of $\mu$ is kept approximately below thresholds as $0.9$ and $0.4$ for the second and third users, respectively.

In Fig. 10, OP curves for FD-NOMA and FD-OMA systems are depicted versus $\sigma^2_{est,SR}$ (while $\sigma^2_{est,l}=0$ and $f_D\tau=0$ for $R-U_l$ link) and $\sigma^2_{est,l}$ (while $\sigma^2_{est,SR}=0$ and $f_D\tau=0$ for $S-R$ link), respectively, in case $n_R=1$, $\mu=0.25$, $SNR=15$ dB and $m_{SR}=m_{RR}=m_{RU}=1$. We set the parameter of FBD effect as $f_D\tau=0.03$ corresponding to the examined link. We observe that OP performance of all users gets worse as the effect of CEE in both hops increases. In addition, a significant performance gain is achieved at all values of CEE effects by increasing the number of antennas at the BS. However, the performance gap between $n_B=3$ and $n_B=2$ results obtained according to CEE effect of the second hop ($\sigma^2_{est,l}$) is much less than the result obtained for the first hop. Therefore, practical impairments in the second hop is more destructive in the performance than those in the first hop. Moreover, the investigated system outperforms FD-OMA for the third user at all values of CEE effects (for both hops) whereas the value of CEE should be kept below a certain threshold such as approximately $\sigma^2_{est,SR}=0.125$ for the second user in case of $n_B=3$ according to results obtained versus $\sigma^2_{est,SR}$.      

\begin{figure}[!t]
	\centering
	\includegraphics[width=0.85\columnwidth]{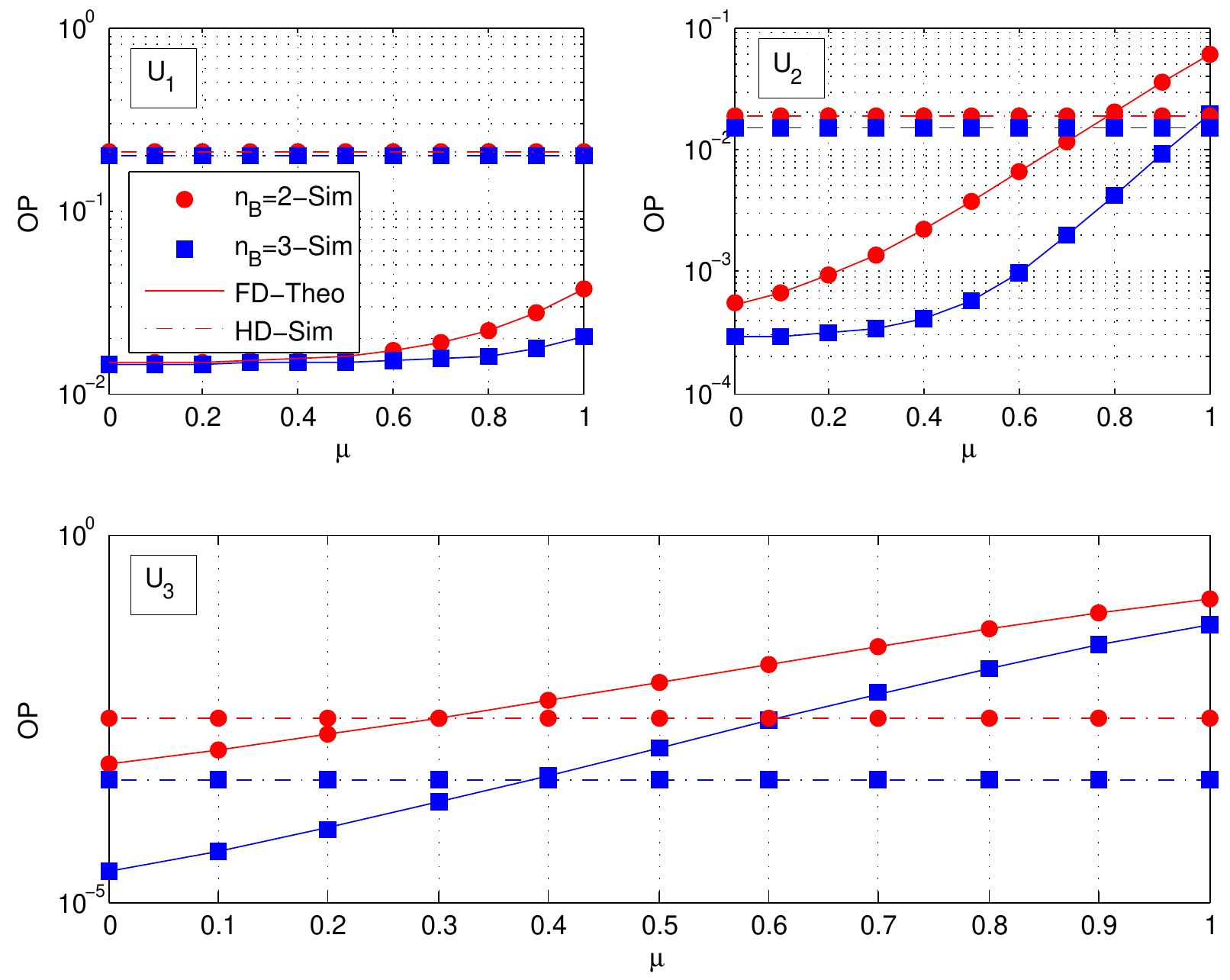}
	\caption{OP comparison of the investigated FD-NOMA system with HD-NOMA counterpart versus $\mu$ in case $n_R=1$, $SNR=15$ dB and $m_{SR}=m_{RR}=m_{RU}=1$ for ideal conditions.}
	\label{fig:7}
	\vspace{-3 mm}
\end{figure}
\begin{figure}[!t]
	\centering
	\includegraphics[width=0.85\columnwidth]{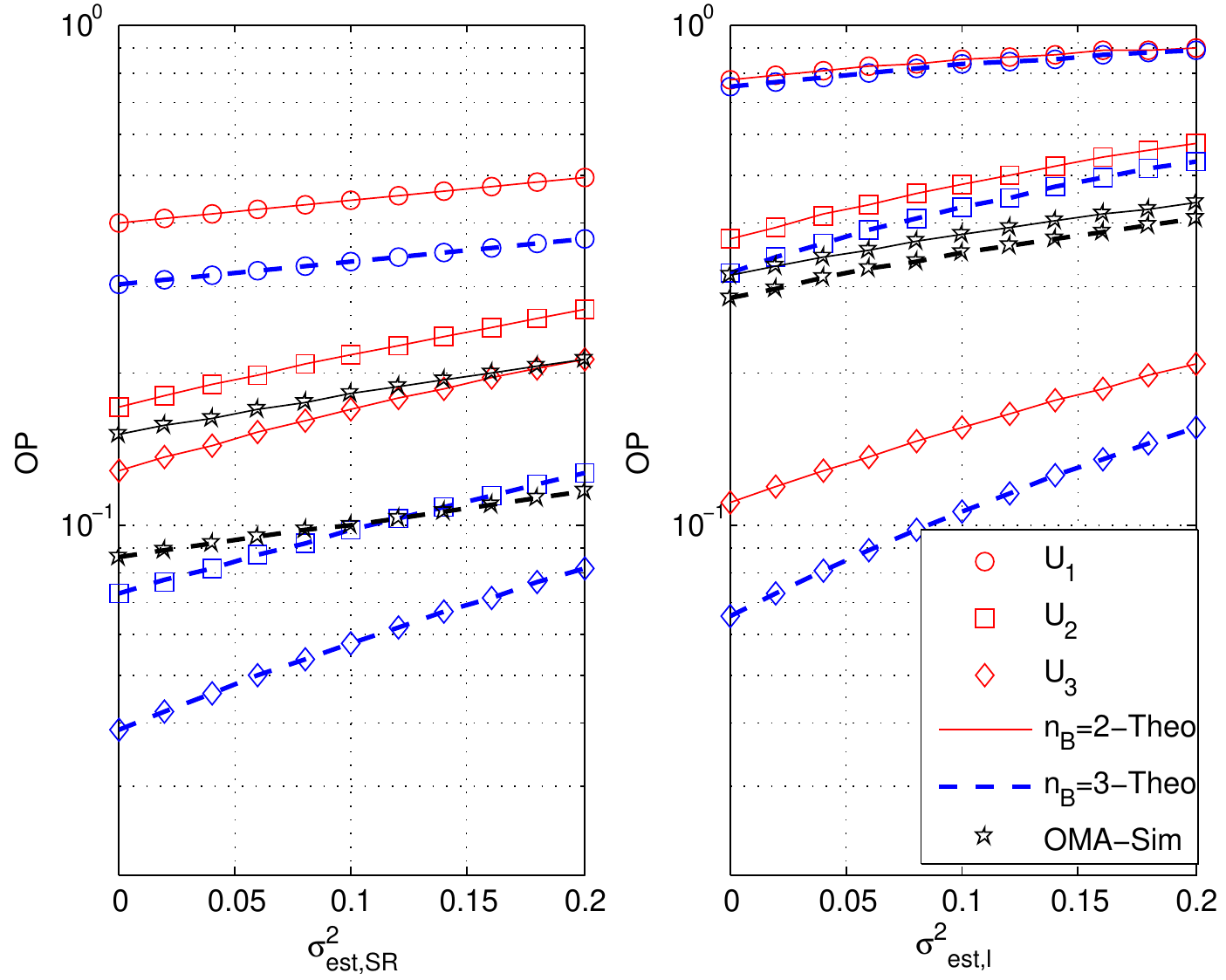}
	\caption{OP comparisons of the investigated FD-NOMA system with FD-OMA counterpart versus $\sigma^2_{est,SR}$ and $\sigma^2_{est,l}$, respectively, in case $n_R=1$, $\mu=0.25$, $SNR=15$ dB and $m_{SR}=m_{RR}=m_{RU}=1$ for ideal conditions.}
	\label{fig:8}
	\vspace{-2 mm}
\end{figure}
\begin{figure}[!t]
	\centering
	\includegraphics[width=0.85\columnwidth]{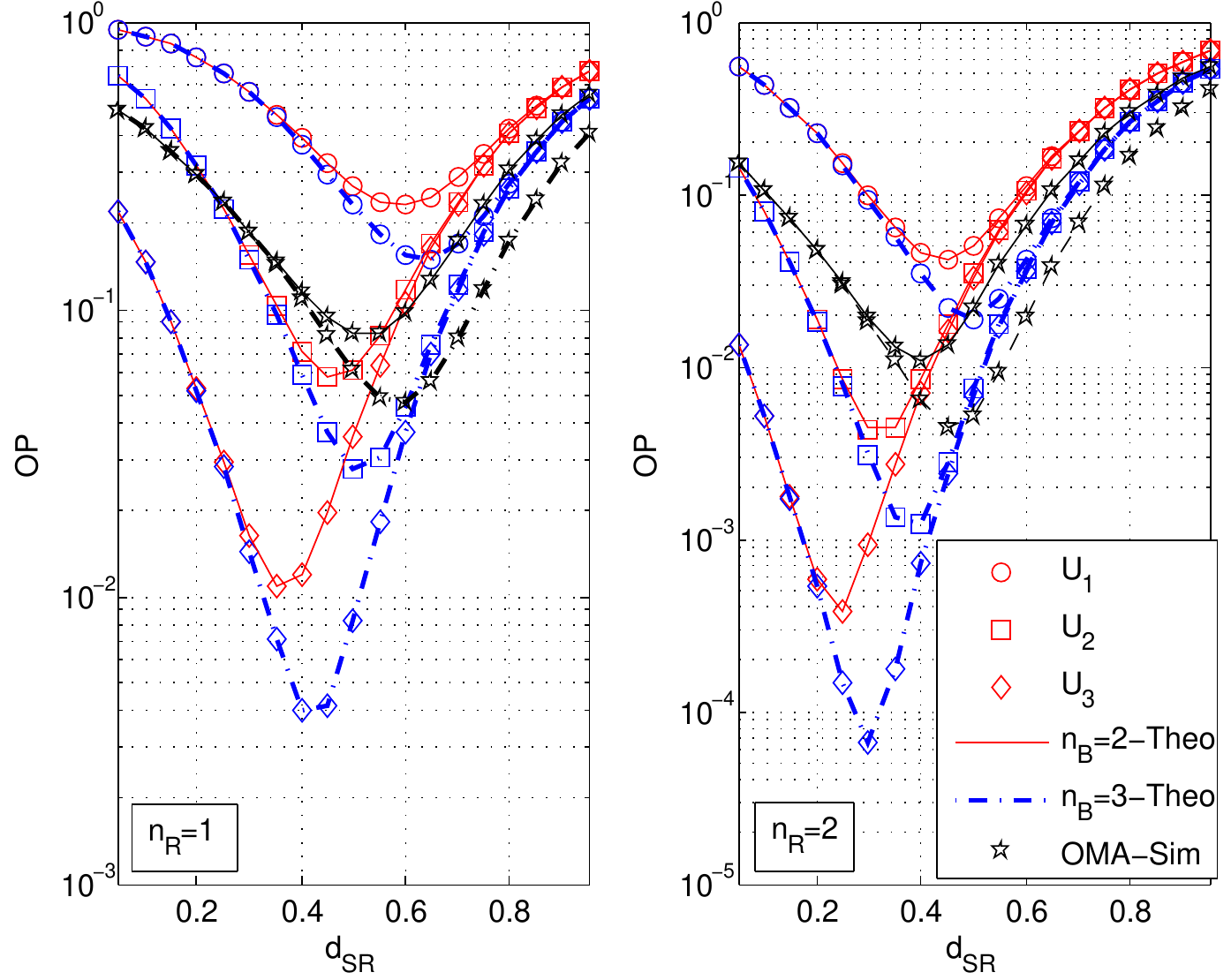}
	\caption{OP comparisons of the investigated FD-NOMA system with FD-OMA counterpart versus $d_{SR}$ in case $\mu=0.25$, $SNR=15$ dB and $m_{SR}=m_{RR}=m_{RU}=1$ for ideal conditions.}
	\label{fig:9}
	\vspace{-2 mm}
\end{figure}
\begin{figure}[!t]
	\centering
	\includegraphics[width=0.85\columnwidth]{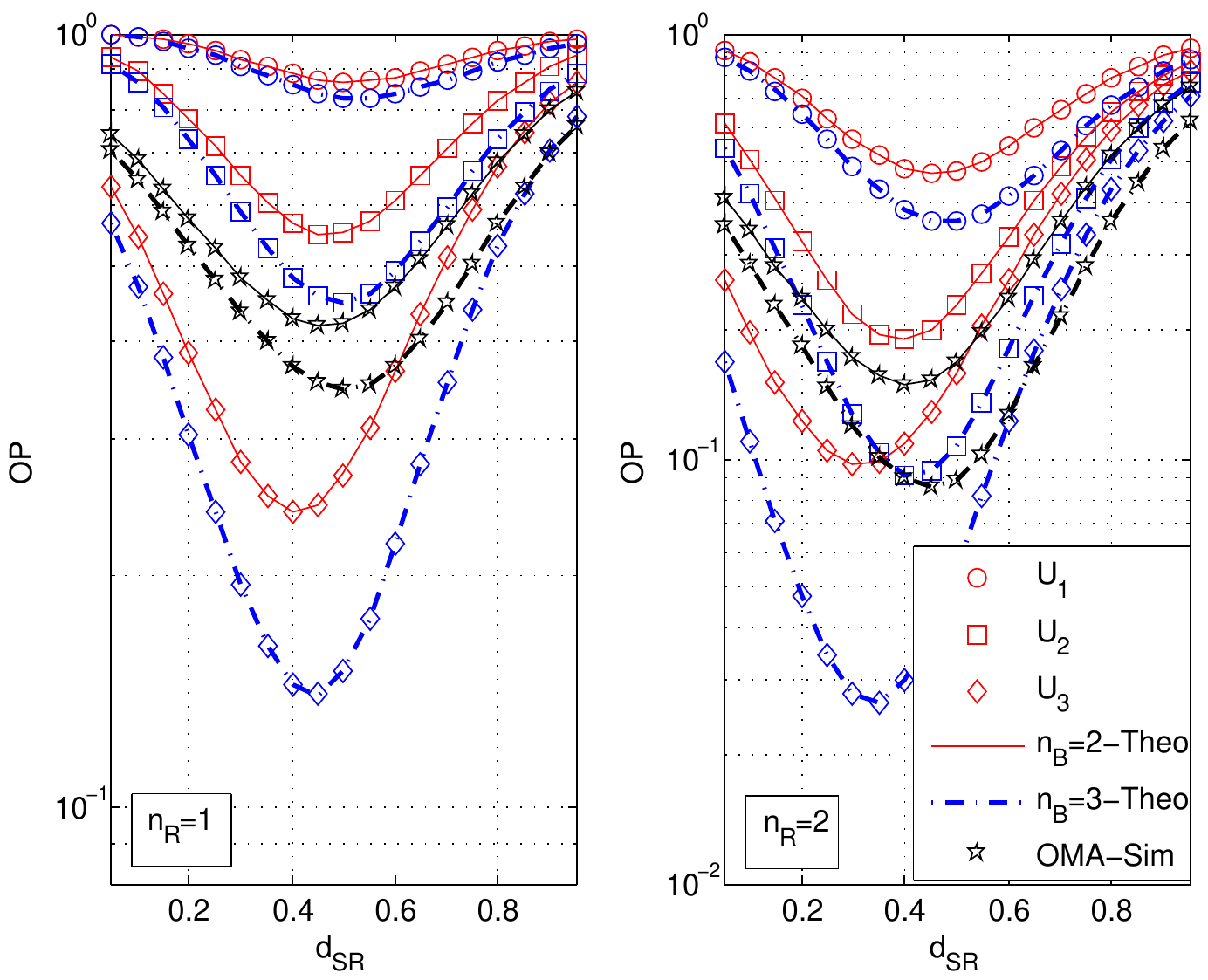}
	\caption{OP comparisons of the investigated FD-NOMA system with FD-OMA counterpart versus $d_{SR}$ in case $\mu=0.25$, $SNR=15$ dB and $m_{SR}=m_{RR}=m_{RU}=1$ in the presence of CEE and FBD.}
	\label{fig:10}
	\vspace{-2 mm}
\end{figure}

Figs. 11 and 12 depict OP comparisons of FD-NOMA and FD-OMA systems versus $d_{SR}$ for ideal and practical conditions, respectively. OP curves are obtained for $\mu=0.25$, $SNR=15$ dB, $m_{SR}=m_{RR}=m_{RU}=1$ and different number of transmit/receive antennas. For practical condition, values of CEE and FBD effects are set as $\sigma^2_{est,SR}=\sigma^2_{est,l}=0.01$ and $f_D\tau=0.03$. The distances between relay and users are determined according to $d_{RU_l}=1-d_{SR}$. From Fig. 11, we observe that the OP of any user is generally minimum when the location of relay is chosen as $d_{SR}<d_{RU_l}$. The main reason for this result is that optimal relay location has a relationship with diversity order and array gain. Therefore, optimal relay location cases can be determined as $d_{SR}<d_{RU_l}$, $d_{SR}>d_{RU_l}$ and $d_{SR}\simeq d_{RU_l}$ depending on relations of $(1-\mu)m_{SR}n_B<m_{RU_l}n_Rl$, $(1-\mu)m_{SR}n_B>m_{RU_l}n_Rl$ and $(1-\mu)m_{SR}n_B=m_{RU_l}n_Rl$, respectively. On the other hand, according to Fig.12, optimum relay locations approach to the middle between the BS and users in the presence of CEE and FBD since the system suffers from zero diversity, therefore only array gain has an impact on the locations. In addition, in order for the proposed FD-NOMA system exhibits better performance than FD-OMA, the relay should be kept close to the BS, especially for the second and third users.

\section{Conclusion}
In this paper, performance of conventional TAS/Alamouti-STBC scheme is investigated in a dual-hop FD AF relaying network based on power-domain downlink multi-user MIMO-NOMA. MRC technique is employed at users and practical impairments CEE and FBD are also taken into account. All performance analyses are conducted over Nakagami-$m$ fading channels and exact OP for any user is derived in single-fold integral form. Simple lower bounds and asymptotic expressions providing information about diversity order and array gain metrics are derived to obtain more meaningful insights. Accuracy of the theoretical expressions are verified through Monte Carlo simulations and SDR-based measurements. According to results, we observe that the better quality of SI cancellation process in the FD relay, the better OP performance gain can be achieved for all users. In addition, the quality of SI cancellation is mostly effective in OP performances of the second and third users. The number of antennas in the second hop is effective in performances of all users while the first hop is quite effective in performances of the second and third users. In case of CEE and FBD effects, the investigated system suffers from an error floor which is also verified by asymptotic analyses, however it outperforms FD-OMA under all CEE effects (which come from both hops) for the third user. Also, performance of all users increases as the channel condition gets better. The investigated system outperforms HD-NOMA counterpart at all values of $\mu$ (specifies the quality of SI cancellation) for the first user while the value of $\mu$ should be kept below a certain threshold for the second and third users, respectively. Moreover, optimal relay location providing minimum OP in the investigated system is observed to have a strong relation with diversity order and array gain in ideal condition while it is only effected by array gain in a practical manner.       

\vspace{-2mm}
\appendices
\section{Proof of Theorem 1}
In order to obtain (\ref{eq:11}), we should firstly determine channel statistics $f_{A}(x)$, $f_{B}^{(l)}(x)$ and $f_{C}(x)$ and then substitute them into (\ref{eq:10}). In order to find the PDF of $f_{A}(x)$, random variable $A$ can be represented as $A=\|\hat{\tilde{\mathbf{h}}}^{\tau}_{SR}\|_F^2=|\hat{h}^{1,\tau}_{SR}|^2+|\hat{h}^{2,\tau}_{SR}|^2=X+Y$. Since channels are distributed as i.i.d. Nakagami-$m$ fading, PDF and CDF of any random variable $\Upsilon=|h^i|^2$ corresponding to any link are expressed as $f_{\Upsilon}(x)=(m/\Omega_{\Upsilon})^m\frac{x^{m-1}}{\Gamma(m)}e^{-xm/\Omega_{\Upsilon}}$ and   $F_{\Upsilon}(x)=1-e^{-xm/\Omega_{\Upsilon}}\sum_{n=0}^{m-1}\frac{(xm/\Omega_{\Upsilon})^n}{n!}$ which follow Gamma distribution. Now, joint PDF of $X$ and $Y$ is written as $f_{XY}(x,y)=n_B(n_B-1)f_X(x)f_Y(y)(F_Y(y))^{n_B-1}$. By using the MGF definition of $\mathcal{M}_{\Upsilon}(s)=\int_{0}^{\infty}e^{-sx}f_{\Upsilon}(x)dx$ and joint PDF written above, the MGF of random variable $A$ can be expressed as 
\setcounter{equation}{21}  
\begin{equation}\label{eq:22}
\begin{split}
\mathcal{M}_{A}(s)=n_B(n_B-1)&\int_{0}^{\infty}\int_{0}^{x}e^{-s(x+y)}f_X(x)f_Y(y) \\
&(F_Y(y))^{n_B-2}dydx. 
\end{split}
\end{equation}
In (\ref{eq:22}), by using binomial expansion \cite[eq.(1.111)]{Gradshteyn} and power series methods \cite[eq.(0.314)]{Gradshteyn}, the CDF component denoted by $\Psi=(F_Y(y))^{n_B-2}$ can be obtained in closed-form as  
\begin{equation}\label{eq:23}
\Psi=\sum\limits_{r=0}^{n_B-2}\sum\limits_{n=0}^{r(m_{SR}-1)}\binom{n_B-2}{r}(-1)^r\beta_n(r,m_{SR})y^ne^{-y\frac{m_{SR}r}{\hat{\Omega}_{SR}}},
\end{equation}
where $\binom{\cdot}{\cdot}$ is binomial coefficient. Here, $\beta_n(r,m_{SR})$ denotes multinomial coefficient consisting of recursive summation and is valid in case of $n\geq 1$ \cite{Mtoka}. Then, by substituting corresponding PDFs and (\ref{eq:23}) into (\ref{eq:22}), MGF of random variable $A$ can be derived as   
\begin{equation}\label{eq:24}
\begin{split}
&\mathcal{M}_{A}(s)=\frac{n_B(n_B-1)}{(\Gamma(m_{SR}))^2}\left(\frac{m_{SR}}{\hat{\Omega}_{SR}} \right)^{2m_{SR}}\sum\limits_{r=0}^{n_B-2}\sum\limits_{n=0}^{r(m_{SR}-1)}\binom{n_B-2}{r} \\
&(-1)^r\beta_n(r,m_{SR})\frac{\Gamma(2m_{SR}+n)}{(m_{SR}+n)}\left(2s+\frac{m_{SR}(2+r)}{\hat{\Omega}_{SR}} \right)^{-2m_{SR}-n} \\
&{}_2\mathcal{F}_1\left(1;2m_{SR}+n;m_{SR}+n+1;\frac{s+\frac{m_{SR}(1+r)}{\hat{\Omega}_{SR}}}{2s+\frac{m_{SR}(2+r)}{\hat{\Omega}_{SR}}} \right),  
\end{split}
\end{equation}
where integral expressions provided in \cite[eq.(3.381.1)]{Gradshteyn} and \cite[eq.(6.455.2)]{Gradshteyn} are used. ${}_2\mathcal{F}_1(\cdot;\cdot;\cdot)$ denotes the Gauss hypergeometric function \cite[eq.(9.14)]{Gradshteyn}, and by using its series representation and inverse Laplace transform property which is defined as $f_{\Upsilon}(x)=\mathcal{L}^{-1}\left\lbrace\mathcal{M}_{\Upsilon}(s)\right\rbrace $, PDF of random variable $A$ can be obtained as follows
\begin{equation}\label{eq:25}
\begin{split}
&f_{A}(x)=\frac{n_B(n_B-1)}{(\Gamma(m_{SR}))^2}\left(\frac{m_{SR}}{\hat{\Omega}_{SR}} \right)^{2m_{SR}}\sum\limits_{r=0}^{n_B-2}\sum\limits_{n=0}^{r(m_{SR}-1)}\sum\limits_{t=0}^{m_{SR}-1} \\
&\binom{n_B-2}{r}\frac{(-1)^r}{t!}2^{-t-n-m_{SR}}\beta_n(r,m_{SR})\Gamma(m_{SR}+n+t) \\
&\sum\limits_{t_1=1}^{T}\sum\limits_{t_2=1}^{\alpha_{t_1}}\frac{\kappa_{t_1t_2}x^{t_2-1}e^{-xs_{t_1}}}{\Gamma(t_2)}. 
\end{split}
\end{equation} 

In order to obtain (\ref{eq:25}), partial fraction decomposition (PFD) method provided in \cite[eq.(2.102)]{Gradshteyn} is applied to (\ref{eq:24}) and then inverse Laplace transform property given in \cite[eq.(2.2.1-1)]{Prudnikov} is used. In (\ref{eq:25}), $\kappa_{t_1t_2}$ denotes the coefficient coming from PFD method. If subindex of the first summation is $r=0$, then $T=1$, $s_1=m_{SR}/\hat{\Omega}_{SR}$ and $\alpha_1=n+2m_{SR}$. On the other hand, if $r\neq 0$, then $T=2$, $s_1=m_{SR}/\hat{\Omega}_{SR}$, $s_2=m_{SR}(2+r)/2\hat{\Omega}_{SR}$, $\alpha_1=-t+m_{SR}$ and $\alpha_2=t+n+m_{SR}$.

Furthermore, PDF and CDF of unordered random variable $B$ can be respectively obtained as $f_{B}(x)=(m_{RU_l}/\hat{\Omega}_{RU_l})^{m_{RU_l}n_R}\frac{x^{m_{RU_l}n_R-1}}{\Gamma(m_{RU_l}n_R)}e^{-xm_{RU_l}/\hat{\Omega}_{RU_l}}$ and   $F_{B}(x)=1-e^{-xm_{RU_l}/\hat{\Omega}_{RU_l}}\sum_{k_1=0}^{m_{RU_l}n_R-1}\frac{(xm_{RU_l}/\hat{\Omega}_{RU_l})^{k_1}}{k_1!}$. Also, $l$th order statistic of random variable $B$ can be represented in terms of PDF as $f_B^{(l)}(x)=Q_l\sum_{k=0}^{L-l}(-1)^k\binom{L-l}{k}f_B(x)(F_B(x))^{l+k-1}$, where $Q_l=L!/(L-l)!(l-1)!$, \cite{Mtoka,JJmen2}. By using these preliminaries, PDF of $l$th order of random variable $B$ can be obtained as
\begin{equation}\label{eq:26}
\begin{split}
&f_B^{(l)}(x)=\sum\limits_{k=0}^{L-l}\sum\limits_{p=0}^{l+k-1}\sum\limits_{k_1=0}^{p(m_{RU_l}n_R-1)}\binom{L-l}{k}\binom{l+k-1}{p}\frac{Q_l(-1)^{k+p}}{\Gamma(m_{RU_l}n_R)} \\ 
&\left( \frac{m_{RU_l}}{\hat{\Omega}_{RU_l}}\right)^{m_{RU_l}n_R}\beta_{k_1}(p,m_{RU_l}n_R)x^{m_{RU_l}n_R+k_1-1}e^{-x(1+p)\frac{m_{RU_l}}{\hat{\Omega}_{RU_l}} }. 
\end{split}
\end{equation}
The PDF of random variable $C$ ($f_C(x)$) has the same expression as $f_{\Upsilon}(x)$. Finally, by substituting the obtained PDFs $f_{A}(x)$, $f_{B}^{(l)}(x)$ and $f_{C}(x)$ into (\ref{eq:10}) and solving the three-fold integral with the help of properties given by \cite[eq.(8.352.4)]{Gradshteyn} and \cite[eq.(3.471.9)]{Gradshteyn}, (\ref{eq:11}) is obtained. Thus, the proof of Theorem 1 is completed. 

\section{Proof of Theorem 2}
The CDF of predefined random variable $W\stackrel{\triangle}{=}\bar{\gamma}A/\left( \bar{\gamma}C+\vartheta_4^{\prime}\right) $ can be expressed as
\begin{equation}\label{eq:27}
\begin{split}
F_W(x)&=Pr\left(A\leq x\left(C+\frac{\vartheta_4^{\prime}}{\bar{\gamma}} \right)  \right) \\
&=1-\int\limits_{y=0}^{\infty}\int\limits_{x=x\left(y+\frac{\vartheta_4^{\prime}}{\bar{\gamma}} \right) }^{\infty}f_A(x)f_C(y)dxdy  
\end{split}
\end{equation}
If we substitute PDFs of $f_A(x)$ and $f_C(x)$ derived in \textit{Appendix A} into (\ref{eq:27}), and solve two-fold integral by using properties provided by \cite[eq.(3.381.3)]{Gradshteyn}, \cite[eq.(3.381.4)]{Gradshteyn} and \cite[eq.(8.352.2)]{Gradshteyn}, closed-form of $F_W(x)$ is obtained as in (\ref{eq:14}). Also, by integrating $f_B^{(l)}(x)$ in (\ref{eq:26}), the CDF of ordered random variable $B$ can be obtained as given in (\ref{eq:15}). 

In case of ideal conditions, by following the same steps, the CDF of $F_W(x)$ can be also derived as in (\ref{eq:16}) while $f_B^{(l)}(x)$ remains the same, and so Theorem 2 is proved. 

\section{Proof of Theorem 3}    
Firstly, asymptotic CDF of $F_{W}^{\infty}(2\varLambda_l^{\dag})$ can be mathematically defined as $F_{W}^{\infty}( 2\varLambda_l^{\dag})=\int_{0}^{\infty}F_A^{\infty}( 2\varLambda_l^{\dag}y)f_C(y)dy$. To obtain CDF $F_A^{\infty}(x)$, asymptotic CDF and PDF expressions related to any random variable $\Upsilon$ defined before should be found. Hence, by using the asymptotic property of incomplete Gamma function $\gamma(v,x\rightarrow 0)\approx x^v/v$ \cite[eq.(45:9:1)]{Oldham}, CDF and PDF can be asymptotically obtained as $F_{\Upsilon}^{\infty}(x)=\frac{(xm/\Omega_{\Upsilon})^m}{\Gamma(m+1)}$ and $f_{\Upsilon}^{\infty}(x)=\frac{(m/\Omega_{\Upsilon})^mx^{m-1}}{\Gamma(m)}$, respectively. Then, if these expressions are substituted into (\ref{eq:22}), MGF of the random variable $A$ can be asymptotically derived as
\begin{equation}\label{eq:28}
\begin{split}
\mathcal{M}_A^{\infty}(s)=&\sum_{t=0}^{m_{SR}-1}\frac{n_B(n_B-1)\Gamma(m_{SR}(n_B-1)+t)}{\Gamma(m_{SR})(\Gamma(m_{SR}+1))^{n_B-2}} \\
&\frac{\left(m_{SR}/\Omega_{SR}\right)^{m_{SR}n_B}}{2^{m_{SR}(n_B-1)+t}t!s^{m_{SR}n_B}}. 
\end{split}
\end{equation}
In order to obtain (\ref{eq:28}), integral properties given in \cite[eq.(3.381.1)]{Gradshteyn}, \cite[eq.(6.455.2)]{Gradshteyn} and series expansion in \cite[eq.(9.14)]{Gradshteyn} are used. Then, by using the inverse Laplace transform property $F_{\Upsilon}(x)=\mathcal{L}^{-1}\left\lbrace\mathcal{M}_{\Upsilon}(s)/s\right\rbrace $, asymptotic CDF $F_A^{\infty}(x)$ can be derived as                    
\begin{equation}\label{eq:29}
\begin{split}
F_A^{\infty}(x)=&\sum_{t=0}^{m_{SR}-1}\frac{n_B(n_B-1)\Gamma(m_{SR}(n_B-1)+t)}{\Gamma(m_{SR})(\Gamma(m_{SR}+1))^{n_B-2}} \\
&\frac{\left(m_{SR}/\Omega_{SR}\right)^{m_{SR}n_B}}{\Gamma(m_{SR}n_B+1)2^{m_{SR}(n_B-1)+t}t!}x^{m_{SR}n_B}. 
\end{split}
\end{equation}
Afterwards, by substituting (\ref{eq:29}) and $f_C(y)$ into the integral representation of $F_{W}^{\infty}(2\varLambda_l^{\dag})$, the result of $F_{W}^{\infty}(2\varLambda_l^{\dag})=\left(\Xi_1\bar{\gamma} \right)^{-(1-\mu)m_{SR}n_B} $ is obtained, where $\Xi_1$ is provided in (\ref{eq:21}) and integral property given in \cite[eq.(3.381.4)]{Gradshteyn} is used. On the other hand, by using the unordered asymptotic CDF $F_B^{\infty}(x)$ (which is the same as $F_{\Upsilon}^{\infty}(x)$), the CDF of ordered random variable $B$ can be asymptotically derived as  
\begin{equation}\label{eq:30}
\begin{split}
F_B^{(l),\infty}(x)&=\binom{L}{l}(F_B^{\infty}(x))^l \\
&=\binom{L}{l}\left(\frac{\left( xm_{RU_l}/\Omega_{RU_l}\right)^{m_{RU_l}n_R} }{\Gamma(m_{RU_l}n_R+1)} \right)^l.
\end{split}
\end{equation}
In order to obtain (\ref{eq:30}), only the lower order terms related to variable $x$ are taken into account. Then, by changing variable $x$ with $(2/\bar{\gamma})\varLambda_l^{\dag}$, the result of $F_B^{(l),\infty}(x)=\left( \Xi_2\bar{\gamma}\right)^{-m_{RU_l}n_Rl} $ is obtained, where $\Xi_2$ is provided in (\ref{eq:21}). Eventually, the proof of Theorem 3 is completed.   




\end{document}